\documentclass[a4paper,11pt]{article}
\usepackage[utf8]{inputenc}
\usepackage[T1]{fontenc}
\usepackage{amsmath}
\usepackage{amsfonts}
\usepackage{amssymb}
\usepackage{nicefrac}
\usepackage{authblk}
\usepackage{fullpage}

\usepackage{multicol}

\usepackage{ifpdf}
\usepackage{url}

\ifpdf
 \usepackage[pdftex]{graphicx}
 \usepackage[pdftex]{color}
\else
 \usepackage{graphicx}
 \usepackage{color}
\fi
\usepackage{wrapfig}
\usepackage{verbatim}
\usepackage{tikz-cd}

\usepackage{tcolorbox}
\definecolor{aliceblue}{rgb}{0.94, 0.97, 1.0}

\usepackage{wrapfig}

\usepackage{natbib}

\newcommand{\rem}[1]{}
\def\N{\mathcal{N}}

\renewenvironment{thebibliography}[1]{
  \begin{oldthebibliography}{#1}
    \setlength{\itemsep}{0.3em}
    \setlength{\parskip}{0em}
}
{
  \end{oldthebibliography}
}

\newcommand{\thirdrevision}[1]{\textcolor{black}{#1}}
\newcommand{\revision}[1]{\textcolor{black}{#1}}

\begin{document}


{\LARGE \bf Opinion Polarization by Learning from Social Feedback}\\
{\large S. Banisch and E. Olbrich} \\
Max Planck Institute for Mathematics in the Sciences, Leipzig, Germany
\vspace{-8pt}
\\
\line(1,0) {454}

\vspace{36pt}


\section*{Abstract}
\small
We explore a new mechanism to explain polarization phenomena in opinion dynamics in which agents evaluate alternative views on the basis of the social feedback obtained on expressing them. High support of the favored opinion in the social environment, is treated as a positive feedback which reinforces the value associated to this opinion. In connected networks of sufficiently high modularity, different groups of agents can form strong convictions of competing opinions. Linking the social feedback process to standard equilibrium concepts we analytically characterize sufficient conditions for the stability of bi-polarization. While previous models have emphasized the polarization effects of deliberative argument-based communication, our model highlights an affective experience-based route to polarization, without assumptions about negative influence or bounded confidence.

\normalsize

\section{Introduction}

\thirdrevision{
The public discourse around political polarization has regained momentum in the recent past.
The fundamental divide between Democrats and Republicans in the United States and the rise of nationalistic voices in the European sphere reflect fundamental differences in attitudes and ideas regarding the >>right<< direction to go and have engaged the interest of investigators in social network science because it is a hard problem to model.
With this paper we aim to identify mechanisms and conditions for the emergence of polarization.
Polarization refers either to a distribution of opinions with multiple local maxima or to the process by which such strong divergences of opinions that divide a population come about \citep{DiMaggio1996have,Bramson2016disambiguation}.
The challenge has been to develop an explanation of polarization with mathematical or computational models of opinion dynamics \citep{French1956formal,Abelson1964mathematical,DeGroot1974reaching,Chatterjee1977towards,Friedkin1990social,Deffuant2000mixing,Hegselmann2002opinion}.
We propose a very basic reinforcement learning mechanism that leads to the emergence and persistence of stable patterns of bi-polarization even if the interaction network that encodes the patterns of social influence is strongly connected.
With this mechanism, the paper provides a parsimonious answer to Abelson's old question of >>what on earth one must assume in order to generate the bimodal outcome of community cleavage studies<< \citep[p. 153]{Abelson1964mathematical}.
In contrast to some previous models of} opinion bi-polarization, our approach does not rely on negative social influence \citep{Mark2003culture,Macy2003polarization,Baldassarri2007dynamics,Flache2011small} or on notions of opinion homophily and bounded confidence \citep{Axelrod1997dissemination,Deffuant2000mixing,Hegselmann2002opinion,Maes2013differentiation,Duggins2017psycologically}, but on a simple reinforcement mechanism that has not yet been explored in the context of opinion and polarization dynamics.

The main idea is that individuals express their opinion about an issue and are sensitive to approval and disapproval by their peers \citep{Homans1974social}.
Agreement leads to a positive experience which strengthens \thirdrevision{attachment to the} expressed opinion.
Conversely, disagreement is assumed to be related to a negative sentiment and decreases \thirdrevision{attachment to the} current opinion.
In a wide range of interaction networks with sufficiently high modularity this reinforcement leads to the formation of different opinion clusters in which agents become collectively more and more \thirdrevision{committed to their cluster's opinion.}
The process we propose can be seen as an abstraction from recent models of polarization that rely on ideas from argument persuasion \citep{Dandekar2013biased,Maes2013differentiation}.
However, our model comes with another connotation as the opinion change process is not assumed to involve argument processing but \thirdrevision{more elementary responses that are} mediated by the positive (negative) experience that agreement (disagreement) brings about.

Although this mechanism is derived as a plausible heuristic motivated by psychological research on \emph{implicit} processes of attitude change \citep{Fazio2001automatic,Fazio2004attitude}, its formalization is highly compatible with a subjective and procedural notion of rationality \citep{Simon1978rationality,Goldthorpe1998rational}.
It gives rise to a reward-driven reinforcement learning scheme that is psychologically plausible yet minimal. 
In the course of the process, agents internalize the expected opinion in their neighborhood and learn to associate values to the different opinion expressions that converge to the payoffs in the corresponding >>opinion game<<.
This means that classical equilibrium concepts can be applied to characterize the stable macroscopic outcomes of the opinion formation process.
This bridge from a rather basic process of opinion formation to a setting where game-theoretic tools become applicable is one of the main 
\thirdrevision{contributions} of the present paper.

\thirdrevision{While} early models of social influence \citep{French1956formal,Abelson1964mathematical,DeGroot1974reaching,Chatterjee1977towards} as well as more recent applications of this paradigm \citep{Friedkin1990social,Friedkin1999choice,Friedkin2016network} allow for analytical treatment, analytical solutions are difficult to obtain if non-linearities \revision{such as bounded confidence \citep{Deffuant2000mixing,Hegselmann2002opinion} or continuous forms of opinion-dependent interaction weights \citep{Maes2013differentiation,Duggins2017psycologically}} are introduced (cf. \cite{Hegselmann2002opinion} and \cite{Flache2011small}).
Consequently, new approaches to the modeling of opinion bi-polarization that generally come with non-linear extensions to produce the desired behavior rely on agent-based simulations to explore the model behavior.
In this paper, we \thirdrevision{also} use simulations to illustrate that opinion bi-polarization is possible -- indeed likely -- by the social feedback account.
One main objective of the computational experiments, however, is to explore and validate the connection from a plausible social feedback mechanism to game-theoretic notions of equilibrium through the use of reinforcement learning.
Once this connection is established, the concept of cohesion as used in social network analysis \citep{Wasserman1994social} and adopted in the theory of games on networks \citep{Morris2000contagion, Jackson2014games} provides a precise structural condition for the stability of bi-polarization in heterogeneous networks.

\thirdrevision{The} incorporation of ideas from reinforcement learning in the context of opinion dynamics bears great potential as a new modeling paradigm.
It differs fundamentally from most previous modeling approaches in conceiving the articulation of an opinion as a communication act that reflects but does not directly correspond to an agent's attitudinal evaluation of an issue.
While, in this paper, the decision of what opinion to express is based on an agent's conviction that its opinion is approved by peers, the framework is general enough to incorporate situational factors as well as strategic considerations that may be involved in opinion statements.
It assumes that agents express their opinions in their social environment and that peers may respond to this communication in different ways.
These responses provide feedback to the sender and sometimes, in some media settings, the sheer number of responses provides a valuable reward.
Implicitly or explicitly, these rewards lead to a re-evaluation of the expression that has triggered the responses which will affect the future behavior of that agent.
This framework, therefore, may shift the explanatory focus from forms of social influence (e.g. strong versus weak or positive versus negative) to the incentives and rewards of opinion expression in different social settings.

The opinion polarization model we devise and analyze throughout this paper 
\thirdrevision{is} a first illustration of this more general program.
We focus on dyadic interaction events constrained by a time-homogeneous network and show that the existence of cohesive subgroups \citep{Wasserman1994social,Morris2000contagion} is sufficient to generate stable bi-polarization even if the subgroups are connected.
\revision{
On the other hand, this simple model shows that very basic social feedback mechanisms may be involved in processes by which an initially moderate population polarizes into two camps which strongly support opposing views and its formulation in terms of reinforcement learning and game theory provides a new account of polarization processes that can be studied using analytical tools.}
With regard to more recent proposals to model bi-polarization, our model comes with a minimal set of individual-level assumptions and does, most notably, not rely on opinion homophily or bounded confidence by which interaction probabilities depend on opinion similarity.
In that sense, our model identifies a previously unseen combination of mechanisms that can explain the emergence and stability of bi-polarization.

There are two basic properties that -- in combination -- lead to the emergence and persistence of bi-polarization in the proposed model.
On the one hand, the reinforcement mechanism gives rise to a group polarization process by which a densely connected group initially inclined into one attitudinal direction becomes more extreme in interaction.
See, for instance, \cite{Sunstein2002law} for a review of ample empirical evidence on this phenomena.
On the other hand, the model exhibits >>gate keeping<< such that opinions do not spread across structural holes \citep{Burt2004structural} in between different communities which prevents a single opinion to spread over the entire network.
We focus on these two properties and their interplay in Section \ref{sec:math}.

In the presentation of our approach we proceed in the following way.
In Section \ref{sec:2} we review the field of opinion dynamics with a special focus on recent approaches to model opinion bi-polarization.
We describe our model in Section \ref{sec:3} and provide all the details on the implementation to allow for the reproduction of the results.
In Section \ref{sec:4} we take a macroscopic and a microscopic perspective to analyze model realizations on random geometric graphs \citep{Penrose2003random,Dall2002random} in order to illustrate how the mechanism leads to stable polarization.
Section \ref{sec:math} is devoted to the mathematical analysis of the model and establishes the fact that the learning scheme approaches the corresponding >>opinion game<< in the course of a simulation.
We show that the model leads to a group polarization process by which opinions become extreme in densely connected groups and that structural holes prevent a single opinion to spread over the entire network.
We then focus on a two-community setting to provide a more specific account of how the interplay of these two properties leads to polarization.
In Section \ref{sec:conclusion} we draw a conclusion on the paper as a whole and discuss the implications of combining opinion dynamics with reinforcement learning for future research.

\section{Modeling Opinion Polarization}
\label{sec:2}


Modeling opinion bi-polarization is recently receiving considerable attention in the opinion dynamics community.
In the last few years, a series of possible interaction mechanisms have been proposed by which a population polarizes and approaches a state with a bi-polar opinion distribution \citep{Macy2003polarization,Baldassarri2007dynamics,Flache2011small,Dandekar2013biased,Maes2013differentiation,Maes2015will,Friedkin2015problem,Duggins2017psycologically} and researchers have started to assess the relevance of these competing explanatory variants through experiments \citep{Moussaid2013social,Chacoma2015opinion,Takacs2016discrepancy}.
The comparison of different models in terms of experimentally-founded microscopic assumptions on the one hand and their capabilities to generate plausible macroscopic outcomes on the other is actually a quite remarkable scientific program as it aims at an empirical justification of rather different links in the explanatory chain \citep{Hedstrom2010causal}.

\thirdrevision{
A major motivation for such a mechanism-based approach to polarization has been the fact that most early models of opinion dynamics lead to either consensus or disagreeing opinions that are not polarized.
These social influence models \citep{French1956formal,Abelson1964mathematical,DeGroot1974reaching,Friedkin1990social} are generally based on mechanisms of iterated weighted averaging by which people’s views on an issue tend to become more similar\footnote{See \citep{Takacs2016discrepancy} for a recent experimental confirmation of opinion assimilation}.
Classic models \citep{French1956formal,Abelson1964mathematical,DeGroot1974reaching} do not entail the possibility to generate persistent bi-polarization in a strongly connected (irreducible) network (cf. \cite{Dandekar2013biased}) even if the opinion space becomes arbitrarily complex \citep{DeGroot1974reaching,Chatterjee1977towards}.
}

\revision{
Within this linear class of models, persistent diversity may be achieved by the introduction of stubborn agents that, to a certain degree, are attached to their initial opinion.
In the models devised by Friedkin, Johnsen and co--workers \citep{Friedkin1999choice,Friedkin2011social,Friedkin2015problem,Parsegov2016novel} this is modeled by an inhomogeneous term in the social influence system that balances the importance assigned to the initial opinion against the effects of the social influence process.}
By defining the patterns of mutual influence and persistence of initial opinions appropriately, this framework has proven quite productive in generating a wide range of macro outcomes such as choice shifts \citep{Friedkin1999choice} and most recently the evolution of belief systems \citep{Parsegov2016novel,Friedkin2016network}.
\revision{If individual susceptibilities depend on the initial opinions even a bimodal opinion distribution may be a stable outcome of a linear averaging process \citep{Friedkin2015problem}.
}

\thirdrevision{We note that} global agreement is also the outcome of most socio-physics models of social dynamics which -- based on an analogy to simple spin systems \citep{Ising1925beitrag} -- often conceive opinions as a binary variable (e.g. >>yes<< versus >>no<<) on which agents agree in interaction (e.g. by copying). 
There is a huge body of literature devoted to the analysis of the statistical properties of systems in which agents connected in a network exchange binary opinions and we refer to \citep{Castellano2009statistical} for a review (but see, for instance, \cite{Frachebourg1996exact, Slanina2003analytical, Sood2005voter, Galam2008sociophysics, Banisch2012son, Gleeson2013binary} as well as \cite{Moran1958random,Kimura1964stepping,Clifford1973model} for predecessors in theoretical biology).
As long as no noise \citep{Carro2016noisy} or anti-conformism \citep{Galam2004contrarian} is added to the system, the opinion imitation mechanism leads to global agreement whenever the interaction network is connected, because under this conditions the dynamics give rise to a random walk with absorbing boundaries \citep{Banisch2014dnc,Banisch2016springer}.

The approach taken in this paper can be seen as a combination of physics-inspired binary opinion dynamics and social influence models.
Agents are allowed to express two alternative opinions (binary choice) but evolve a private evaluation of these alternatives that takes continuous values.
Despite the fact that this combination can account for the emergence and persistence of bi-polarization even if the interaction network consists of one connected component, it also leads to microscopically more plausible opinion change dynamics with respect to socio-physics models as agents do not constantly switch back and forth between the two opinions but may go through phases of confidence and indecision (see Fig. \ref{fig:individuals01}).

The first generation of previous opinion dynamics models which try to address the issue of stable opinion plurality are based on the continuous social influence models and include further constraints, most prominently, assumptions about value or \emph{opinion homophily} \citep{McPherson2001birds}.
In the opinion dynamics context this type of homophily refers to the fact that similar views may lead to attraction and an increasing likeliness of interaction.
Paired with positive social influence it describes a process of repeated influence events by which >>similarity leads to interaction and interaction leads to still more similarity<< \citep[109]{Banisch2010acs}.
In the classical models with bounded confidence \citep{Hegselmann2002opinion,Deffuant2000mixing} this is implemented via a threshold mechanism that switches off influence if the opinion discrepancy exceeds a certain threshold. 
In relation to linear social influence models, this effectively means that the matrix of social influences $A$ is a function of the opinions that agents hold which may disintegrate into disconnected groups of influence due to the threshold mechanism.
This procedural disintegration into disconnected influence groups is necessary for persistent disagreement in this class of models.

Also the influential model by \cite{Axelrod1997dissemination} achieves global polarization by mechanisms of local convergence due to an homophily mechanism that disables mutual influence if the individuals become too different.
In his multi-dimensional model, the interaction probability depends on the opinion overlap (i.e. the number of traits the two agents already share) and vanishes if agents are unaligned on all dimensions.
Unlike in other models with a multidimensional representation of opinions \citep{Lorenz2007continuous,Banisch2010acs,Maes2013differentiation}, the Axelrod model relies on a spatial embedding \revision{in form of a lattice} that shares certain similarities with the random geometric graph model we use in Section \ref{sec:4} to illustrate our results.
\revision{This model gives rise to a number of spatial clusters of agents with aligned opinions and these configurations of homogeneous regions become stable as the overlap across regions and consequently the probability of cross--cutting assimilation vanishes.
When run on a spatial topology, our model also gives rise to spatial clusters of opinions, but the interaction probabilities are not affected by this such that agents at the interface between different clusters are still exposed to non-confirming feedback.
The social feedback process stabilizes because the locally prevailing opinion is more often reinforced in the interaction with peers.
Therefore}, while cultural drift or communication errors implemented as small random perturbations are >>eroding the borders<< \citep[p.907]{Centola2007homophily} between culturally stable regions in the cultural dissemination model \citep{Klemm2003global} , the interfaces between different opinion clusters are stable with respect to noise in \revision{the} social feedback model where such perturbations occur naturally through exploration.

Bounded confidence can lead to a stable co-existence of a plurality of opinions in a population. 
However, unless >>extreme<< agents are artificially inserted \citep{Deffuant2002how}, bounded confidence models cannot drive a population into the extremes or lead to the emergence of antagonistic opinion groups as the opinion averaging procedure always locally reduces diversity.
It crucially depends on the initial opinion diversity whether or not multiple opinions survive and the model will always end up with final opinions that are more moderate than the initial extremes.

The recent approaches to modeling polarization dynamics introduce new mechanisms of opinion exchange and influence in addition to positive influence and homophily.
They have been classified into models that rely on the assumption of \emph{negative social influence} and models that draw upon ideas from \emph{persuasion theory} (cf. \cite{Maes2015will}).
\thirdrevision{An alternative introduced in the physics literature on opinion dynamics differentiates between a continuous internal opinion and discrete opinion expression \citep{Martins2008continuous}.}
We will briefly describe all three variants here.

The first branch of models seeks an explanation of polarization patterns by assuming a negative social influence in the interaction of distant agents such that the encounter and communication of two agents with very different views leads them to adopt even more distant positions in the opinion space \citep{Mark2003culture,Macy2003polarization,Baldassarri2007dynamics,Flache2011small}.
Despite the possibility to generate stable polarization patterns driving the population into two maximally extreme clusters, a recent experimental study found no indication for negative influence of this kind \citep{Takacs2016discrepancy}.\footnote{The experiment was indicative of a negative influence only when people held very similar opinions, which can be seen as a tendency to individualization. Assumptions in line with this finding have been investigated in, for instance, \cite{Banisch2010wcss} and \cite{Maes2010individualization}.} 
It is noteworthy that the combination of positive and negative social influence can lead to polarization but that this feature disappears in the presence of opinion homophily \citep{Maes2015will}.

The second type of models capable of explaining polarization is based on psychological work on persuasion and attitude change \citep{Fishbein1963investigation,Lord1979biased,Petty1981personal,Ajzen2001nature}.
Persuasion models generally assume that communication partners exchange arguments about the object on which an attitude is formed and that new arguments are learned from an interaction partner if they are in support of an agents view.
The model by \cite{Maes2013differentiation} is based on an explicit representation of attitudes borrowed from expectancy-value theory which treats an attitude as a weighted combination of a set of pro- and con-arguments regarding certain aspects of the issue under discussion (see also \cite{Urbig2005dynamics} for an early related model).
Their so-called >>argument-communication theory of bi-polarization<< posits a mechanism for the co-evolution of arguments and the associated weights with the latter encoding a limited capacity to process all arguments at once.
That is, only a subset of arguments is considered to be relevant and recently discussed issues play a more important role.
At each time step, two agents are chosen at random with a probability that is proportional to the similarity of their attitudes (opinion homophily).
In the interaction, the first agent adopts a randomly chosen argument from its interaction partner (positive social influence).
The larger homophily, the more likely an agent with a similar opinion will be selected as interaction partner. 
As the similarity in attitudes of both agents may come about by different subsets of arguments (those that are currently relevant), the focal agent is likely to receive an argument in favor of his own attitude from that partner.
As a result, the model leads to a process where agents with a similar attitude mutually reinforce that attitude by the exchange of supportive arguments.\footnote{It is worth noticing that there is a branch in theoretical biology which proposed very similar mechanisms to model sympatric speciation (bimodal distribution of phenotypes) by assortative mating (individuals preferentially mate with similar individuals). See \citep{Kondrashov1998origin} and references therein. Some aspects of the relation between these models and models of opinion dynamics have been discussed in \citep[Chapter 8]{Banisch2016springer}.}

The model proposed in \cite{Maes2013differentiation} operationalizes a deliberative argument-based route to polarization in line with the persuasive argument theory described in the seminal paper on group polarization by \cite{Sunstein2002law}.
Sunstein reviews experimental evidence on group polarization and choice shifts through group discussions and debates on its political and institutional implications.
His argument is that groups inclined to a certain attitudinal direction rely on a limited argument pool which is biased into the respective direction as there is a disproportionate number of supporting claims.
As arguments are exchanged the group members acquire new arguments that tend to speak even more in favor of the initial direction.
In the Mäs/Flache model such biased and limited argument pools come about by an increased interaction probability of agents which already hold similar attitudes leading to enclaves of individuals with similar and every time more extreme inclinations.
It is worth noting that such a persuasive argument account is also in agreement with the functional approach to argumentative reasoning put forth by \cite{Mercier2011humans}, who, however, put more weight on the intuitive sources of attitudes and supporting arguments.
The learning process described in the present paper is potentially more related to such an account.
In particular, it resonates well with the paradigm of automatic (implicit) attitude activation and the learning of evaluative associations put forth by Fazio and colleagues \citep{Fazio2001automatic,Fazio2004attitude}.

Another persuasion-based proposal to the modeling of bi-polarization has been made by \cite{Dandekar2013biased} who base their argument on the work of \cite{Lord1979biased} on \emph{biased assimilation}.
The principle idea is that people who are very convinced of their view tend >>to accept "confirming" evidence at face value while subjecting "disconfirming" evidence to critical evaluation<< (p. 2098).
This is sometimes also referred to as confirmation bias.
Unlike the model described above, \cite{Dandekar2013biased} take an abstract perspective in terms of opinion representation and include biased argument evaluation into the classic repeated averaging model \citep{DeGroot1974reaching} which operates on a single continuous opinion dimension ($o_i \in [0,1]$).
A bias function is introduced that decides about strength and direction of opinion change as a function of the current conviction of one or the other view (the extremes $0$ and $1$ are interpreted as two opposing opinions on an issue).
The model contains DeGroot's averaging process as a limiting case and \cite{Dandekar2013biased} shows that there is a critical bias level at which the process becomes polarizing.
Noteworthy, polarization may occur even in the absence of opinion homophily.

A third mechanism capable of producing a bimodal opinion distribution has been proposed by \cite{Martins2008continuous}.
The model originated from physics-inspired models where agents face binary choices and update their opinion by imitation of their neighbors which generally leads to all agents holding the same opinion.
The main idea of \cite{Martins2008continuous} is to introduce an internal opinion that encodes how many encounters are needed for an agent to change opinion.
That is, whenever an agent observes another one expressing the same opinion, the internal opinion is enforced into the respective direction such that it becomes still harder for the agent to switch.
These internal opinions -- also referred to as inflexibility \citep{Martins2013building} -- may polarize in the sense that two opposing camps of more and more inflexible or extreme supporters of the different options emerge.

The mechanism this paper adds to the polarization literature is related to this last approach by differentiating between an internal evaluation of different discrete options of opinion expression.
Like in persuasion models, opinions are reinforced by encountering agents with similar views, but  in our model this >>reinforcement in agreement<< is mediated by a very different psychological process.
Namely, reinforcement or respectively a weakening of support of their expressed view is not obtained by a costly process of argument persuasion but rather by the positive (negative) experience that agreement (disagreement) brings about.
Agents form their opinions on the basis of the social feedback they obtain by expressing them.
The learning scheme employed in the model is in line with the psychological theory of \cite{Fazio2004attitude} which views attitudes as evaluative associations of varying strength mediated through positive and negative experience.
Recent neuro-physiological studies provide support for such evaluative mechanisms in social interaction \citep[see concluding section]{Campbell2010opinion,Ruff2014neurobiology}.
Moreover, rooted in reinforcement learning our model is amenable to game--theoretic considerations which has recently been proven very productive in relating agent-based models to an economic model of inter-temporal optimization \citep{Banisch2017coconut}.
Here (in Section \ref{sec:math}) we link to the developed body of literature on games on networks \citep[and references therein]{Jackson2014games} and derive analytical results for the stability of polarization.

\section{Model Description}
\label{sec:3}

\subsection{Theoretical Model}

Suppose there are two opinions that agents can adopt and express.
We denote these alternative options as $o_i$, $i$ being an agent index, and set $o_i \in \{-1,1\}$ for further convenience.
In the opinion model we put forth here, an agent (say $i$) is chosen at random and expresses his current opinion $o_i$ to a randomly chosen neighbor $j$.
That is, a first agent $i$ is chosen uniformly from the set of all agents and the second agent $j$ is sampled out of the set of $i$'s neighbors.
Agent $i$ expresses its opinion $o_i$ to agent $j$ and 
this agent responds to $i$'s expression with approval or disapproval (agreement or disagreement) depending on her current opinion $o_j$.

We further assume that agents become more convinced of an opinion if it is approved by their interaction partners and that their conviction in an expressed opinion is challenged if others disagree.
This is accounted for by two real-valued terms per agent -- $Q_i(1)$ and $Q_i(-1)$ -- that capture how well the expression of $1$ and $-1$ respectively is perceived by peers in an agent's social environment.
That is, the $Q_i(o)$ represent an internal evaluation of the different options based on the social response the agent obtains on expressing them.
These values are updated as 
\begin{equation}
Q_i(o) \leftarrow 
\begin{cases}
(1-\alpha) Q_i(o) + \alpha r_i  : \text{ if } o = expression\\
Q_i(o): \text{ else}.
\end{cases}
\label{eq:QUpdate}
\end{equation}
with 
\begin{equation}
r_i = o_i o_j
\label{eq:reward}
\end{equation}
leading to a positive feedback for $o_i = o_j$ and to a negative one if $o_i \neq o_j$.
The parameter $\alpha$ is referred to as learning rate (see below) and governs the magnitude of change.
In the context of social influence opinion dynamics $\alpha$ can be seen as a susceptibility of the agents to revise its opinion evaluation on the basis of social feedback obtained on expressing it.
If not otherwise stated, it is set to $\alpha = 0.05$ for the simulations performed in this paper.

On expressing their current opinion agents thus receive affirmative or non-confirming response depending on the current opinion in their neighborhood.
Agreement signals approval and leads to a positive experience ($r_i = 1$) by which the evaluation $Q_i(o)$ of the respective expressed opinion increases.
Conversely, a disconfirming response gives rise to a negative feedback ($r_i = -1$) and decreases the value associated to the respective opinion.
We shall therefore interpret the values $Q_i(o)$ as associated to the two opinions as a strength with which the respective view is supported by $i$ or likewise as $i$'s conviction regarding the two alternatives.
In particular, in the case where only two competing opinions may be expressed, the difference $\Delta Q_i = Q_i(1)-Q_i(-1)$ can be interpreted as a conviction that one opinion is more favorable.
Consequently, we assume that when asked to articulate their current opinion or to respond to such an articulation, agents choose to express that option which he more strongly supports at the current time step.
In other words, $o_i = -1$ if $Q_i(-1) > Q_i(1)$ and $o_i = 1$ if $Q_i(-1) < Q_i(1)$ or, more generally:
\begin{equation}
o_i = \arg \max_o Q_i(o).
\label{eq:selectionMax}
\end{equation}

However, we assume that agents generally tend to follow (\ref{eq:selectionMax}) in their expression choice, but deviate from this scheme with a small probability $\epsilon = 0.1$.
The mathematical reason for this is that we make sure, in this way, that both opinion options are tested by the agents so that both $Q$-values are actualized from time to time; for this reason $\epsilon$ is usually referred to as exploration rate.
On the other hand, exploration seems also plausible especially when the difference in conviction is small\footnote{In fact, this would motivate the so-called softmax action selection \citep{Sutton1998reinforcement} which assigns equal probability to the two options when there $Q$ values are equal and gradually favors the option with larger $Q$ the more they differ. We have tested this alternative and found no qualitative impact on the behavior of the model.}.
Moreover, one may argue that the noise introduced by $\epsilon$ accounts for certain issues that might occur in communication such as misunderstanding or misinterpretation of an articulation.

Readers familiar with game theory and reinforcement learning will have realized that this model set-up casts opinion dynamics as a game played repeatedly on a network in which agents learn the best response by a simple form of independent Q-learning \citep{Sutton1998reinforcement,Busoniu2008comprehensive}.
In this context, opinion expression should be seen as an action that leads to a certain reward $r$ and the (state-less) Q-values associated to the two possible actions (i.e. $o_i \in \{-1,1\}$) are updated based on this reward signal.
In fact, we will make use of this analogy in the sequel to mathematically characterize a series of stylized situations that will be helpful to provide an overall picture of the model behavior.

We also note that the model behavior is essentially determined by the way in which the reward system is designed.
In this basic model agents learn to avoid dissonance and play a coordination game, but we envision that different more complex and possibly heterogeneous rewards are a propelling ingredient that deserves further exploration.
The main purpose of this paper, however, is to introduce this type of opinion game in the context of opinion models and polarization dynamics in particular, and to demonstrate that opinion polarization can result from very few relatively mild basic assumptions:
\begin{enumerate}
\item 
agreement is positively, disagreement negatively experienced, 
\item
these experiences drive opinion conviction ($\Delta Q_i$) and expression ($o_i$), 
\item
the probability of interaction is structured. 
\end{enumerate}

While the first two assumptions are realized through rewards and the update scheme as specified by (\ref{eq:QUpdate}) - (\ref{eq:selectionMax}), the third aspect is included in the model by an interaction network that determines the probability with which pairs of agents engage in communication with one another.
In order to illustrate the model's capability to generate stable bi-polarization in connected networks with a direct or indirect path between all pairs of agents, 
we generate a random geometric network \citep{Penrose2003random,Dall2002random} to define the neighborhood structure.
According to that network model, $N$ agents are assigned a random position in the unit plane $(x_i,y_i) \in [0,1]\times[0,1]$ and a link is established whenever the distance between two agents is below a threshold $r$.
Notice that the average degree of the network depends on both the number of agents and the threshold $r$ and can be approximated as $\pi N r^2$ for large $N$ \citep{Penrose2003random}.
Consequently, the network density is proportional to $\pi r^2$.
A further question typically addressed in the analysis of random graph models and of special interest to the present discussion concerns the critical threshold $r_c$ above which a giant connected component is formed.
In two dimensions, this threshold has been found by \cite{Dall2002random} to scale as $r_c = 4.52 / N$ with the system size.
Notice finally that random geometric graphs are a special type of random graphs that differ from the latter especially regarding clustering \citep{Dall2002random}.

This graph model is very simple, comes with only a single parameter and is well-suited for the purpose of illustration as it naturally embeds into two dimensions.
In fact, spatial random graphs can be seen as an intermediate between the regular lattice and the classical  Erdős-Rényi random graph model which have both been used frequently in previous simulation studies.
We do not further specify what the similarity in the unit square accounts for but remark that besides spatial proximity it may also mimic social proximity patterns as they come about due to homophily regarding age, status or socio-economic situation \citep{Lazarsfeld1954friendship,McPherson2001birds}.
Notice, again, that contrary to most existing models of opinion dynamics and especially those that aim at explaining bi-polarization, our model comes without any assumption about homophily regarding opinions.
One of the main purposes of this paper is to show that static patterns of interaction and influence as caricatured by random spatial networks are sufficient for bi-polarization to emerge even if indirect influence patterns expand over the entire population.

\subsection{Implementation Details}

\begin{figure}[h]
\centering
\includegraphics[width=0.9\linewidth]{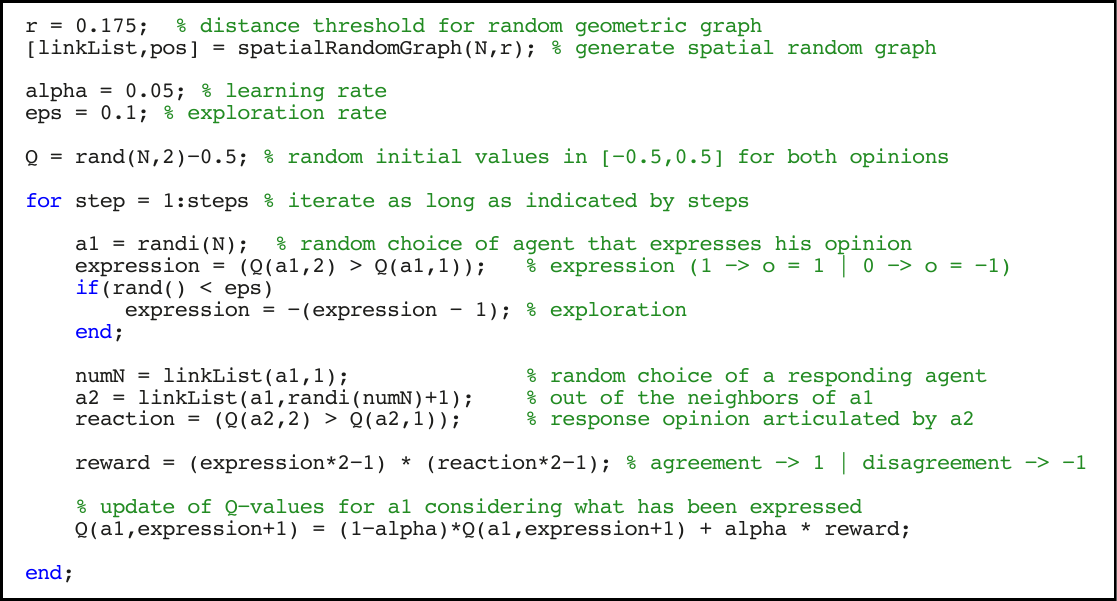}
\caption{Runable MatLab-Code of the model.}
\label{fig:CodeOV}
\end{figure}

\begin{figure}[h]
\centering
\includegraphics[width=0.9\linewidth]{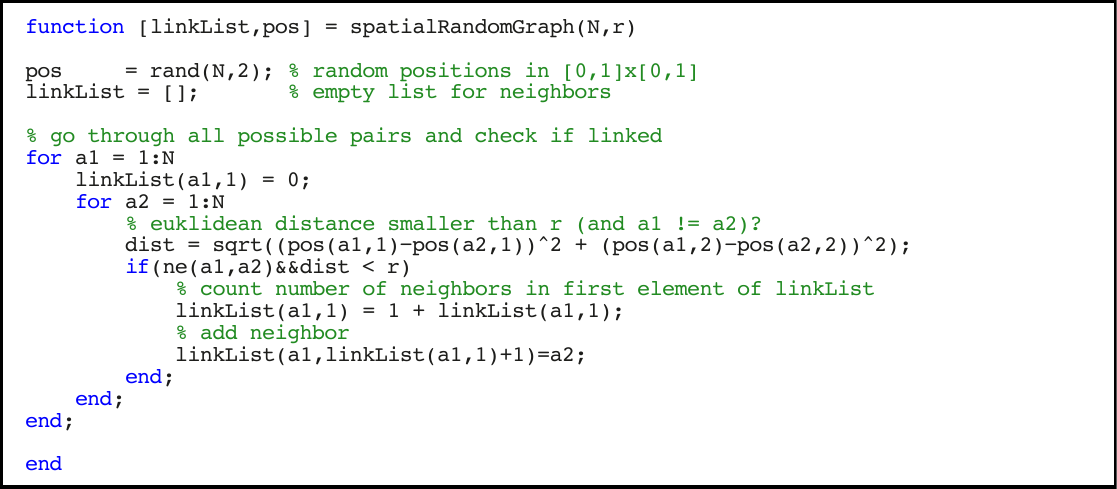}
\caption{Runable MatLab-Code of the random graph model.}
\label{fig:CodeRGM}
\end{figure}

In order to facilitate the reproduction of results and give readers the possibility to make their own explorations with the model, we provide all the details of the model in form of the two code snippets shown in Figure \ref{fig:CodeOV} and \ref{fig:CodeRGM}.
Noteworthy, the model is simple enough to be implemented in a few lines in MatLab (R2013a for Mac) and is executable as it is presented ($N$ and the number of iterations \verb|steps| have to be specified).
Figure \ref{fig:CodeOV} presents the model initialization and the iteration loop, Figure \ref{fig:CodeRGM} the random geometric graph model.
Notice that for the computation-intensive experiments the model was also implemented in C++.

In addition, two online implementations of the model are available under
\begin{itemize}
\item \verb|www.universecity.de/demos/OpinionValuesSmall.html|
\item \verb|www.universecity.de/demos/OpinionValuesBig.html|
\end{itemize}
The former is a model with $N=100$ agents which is also described in the next section.
The latter is an implementation with $N = 10000$ agents where the network is not visualized for performance reasons.
Notice that running the models requires a Browser with WebGL support.

\section{Model Behavior on Random Geometric Graphs}
\label{sec:4}

The aim of this section is to provide an intuition about the model behavior.
We will look in some detail on a realization of the model on a random geometric graph.
We have used a network parameter significantly above the critical value $r > r_c$ in order to generate a graph with a single connected component to highlight that the social feedback model goes beyond social influence models in its capability to generate stable opinion bi-polarization in connected graphs.
We will consider individual trajectories as well as some macroscopic indicators and show that polarization patterns emerge in these networks.

\subsection{Simulation Setting}

As described above we generate a random spatial network to define the neighborhood structure of the agents.
$N$ agents are assigned random position in the unit plane $(x_i,y_i) \in [0,1]\times[0,1]$ and a link is established whenever the distance between two agents is below a threshold $r$.
The agent population is initialized by setting the initial values $Q_i(1),Q_i(-1)$ at random according to a uniform distribution within $[0,1]$.
This defines the initial opinions (i.e. what the agents express at $t=0$).
Due to the initialization of the $Q$--values, on average, half of the population will initially hold opinion $1$ and the other half opinion $-1$.

The exemplary model realization discussed in this section is done with $N = 100$ agents.
The threshold for the random geometric graph model is $r = 0.175$ and the resulting network is shown in Fig. \ref{fig:BS01}.
The learning rate is $\alpha = 0.05$ and the exploration rate $\epsilon = 0.1$.

\subsection{Macroscopic and Microscopic Dynamics}
\label{sec:simulation1}

\begin{figure}[ht]
\centering
\includegraphics[width=0.99\linewidth]{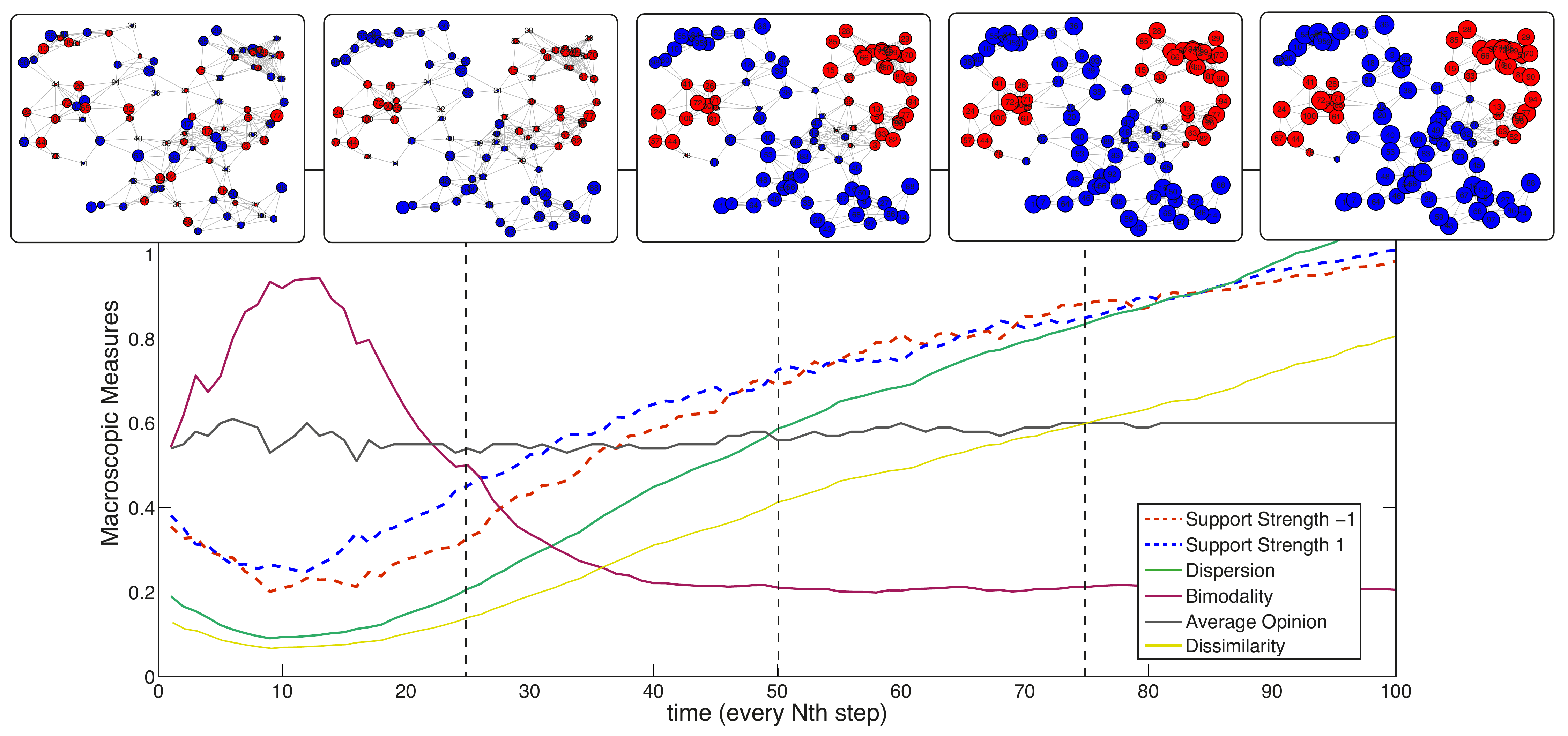}
\caption{Time evolution of the model. The plot shows 10000 iterations for 100 agents (on average 100 expressions by each agent) along with the system state at $t=0, 2500, 5000, 7500, 10000$. The size of the nodes represents their support level and the color which opinion is favored. The average opinion represents the fraction of agents that support $1$ and the support strength is the value difference averaged over the respective sets of supporters. In the definition of the polarization dispersion and bimodality measures we follow \cite{DiMaggio1996have}.}
\label{fig:BS01}
\end{figure}

An example run of the model with $N = 100$ agents is shown in Fig. \ref{fig:BS01}.
It shows for 10000 simulation steps the microscopic system configuration at times zero, 2500, 5000, 7500 and 10000 along with different macroscopic observables that measure the amount of polarization in a population.


Just as polarization mechanisms, also the empirical characterization of polarization patterns is recently receiving some attention and we refer to \cite{Bramson2016disambiguation} for an accessible overview of different measures and a conceptual discussion of the different aspects of polarization they account for.
For the purposes of this paper, we follow \cite{DiMaggio1996have} in the definition of polarization measures and define \emph{dispersion} as the variance $\sigma^2$ over the distribution of convictions $\Delta Q_i = Q_i(1)-Q_i(-1)$ and \emph{bimodality} by its kurtosis
\begin{equation}
\kappa = \frac{ \frac{1}{N} \sum\limits_{i=1}^{N} (\Delta Q_i - \overline{\Delta Q})^4 }{\sigma^4}-3.
\label{eq:kappa}
\end{equation}
Notice that the kurtosis is most often interpreted as a measure of outliers with $\kappa = 0$ for the normal distribution.
Borrowing the interpretation from \cite{DiMaggio1996have}\footnote{The reader is referred to their paper for some example distributions and the respective kurtosis values.}, p. 694-696, positive kurtosis indicates a very peaked consensus distribution whereas it becomes negative for flat and even more so for bimodal distributions reaching $\kappa = -2$ in the two-peaked case.
For the sake of visualizing bimodality in the same interval as the other measures, Fig.\ref{fig:BS01} therefore shows bimodality transformed as $(\kappa + 2)/2$ such that a value of zero indicates complete bimodularity and a value of one no deviation from the normal distribution.

In addition to the measures used by \cite{DiMaggio1996have}, the polarization measure introduced in \cite{Flache2011small} which we refer to as \emph{dissimilarity} is shown.
Dissimilarity is defined as the standard deviation of the distribution of opinion distances between all pairs of agents and is zero for consensus and one for the case of the equally sized groups at the extremes \citep[p. 8]{Maes2013differentiation}.
Moreover, the \emph{average opinion} (fraction of agents that express $o_i=1$) and the \emph{average strength of support} (average over the absolute values of $\Delta Q_i$) within the two different groups of supporters is shown over time.

Initially, due to the random initialization of values, approximately one half of the population supports $expression$ 1 and -1 is supported by the other half. 
Notice that these fractions do not change a lot during this simulation run even if single agents do change their opinion.
The left-most network shows that supporters are initially distributed without a particular spatial organization.
The individual convictions $\Delta Q_i$ accounting for the strength with which the respective alternative is supported (size of the nodes in Fig. \ref{fig:BS01}) is at a relatively low level for both opinions.
The initial average strength of support (dashed curves) -- that is, the average absolute value over convictions of supporters of 1 and -1 respectively -- is around $0.4$ for both alternatives.\footnote{Notice that the theoretically expected value of this measure is 1/3 if both Q-values are drawn from a uniform distribution in $[0,1]$ and that the value of almost 0.4 is the outcome for the randomly drawn initial condition used in this simulation.} 

The strength of support is decreasing during an initial period of alignment because due to the random initial configuration agents meet on average with an equal number of agents from both camps such that no opinion is clearly favored and the $\Delta Q_i$ tend to zero.
However, this initial period results in a strong spatial organization into opinion regions that is clearly visible at step 2500.
This spatial distribution of opinions remains rather stable in the subsequent steps.
While strength of support is still small for most agents at $t=2500$ (approximately at the initial level), some clusters have emerged in which one opinion is more strongly supported after 5000 steps.
In particular, we observe two  regions in which agents strongly support 1 (blue) which are connected by a bend of agents that moderately support 1.
We also observe two disconnected clusters where -1 (red) is supported more and more strongly.
However, decisive changes may still occur at the interfaces between regions in which different opinions are supported and some clusters may be invaded in a long transient.
Despite support strength remaining smaller in those areas, on average, support strength increases for both opinions.

This evolution is also captured by the different polarization measures.
First, dispersion and dissimilarity behave very similarly and show an initial decrease in polarization followed by a steady increase once the opinion clusters have formed.
Kurtosis -- relating to the bimodality of a distribution -- behaves differently.
Namely, there is an initial increase from a moderate value towards a value close to one capturing the fact that the initially uniform distribution approaches a normal one with the $\Delta Q_i$ approaching zero during the very first period.
The first peak in the measure is therefore indicative of a considerable reduction in polarization.
However, as with the other measures, it indicates that polarization increases to the initial level after approximately 2500 steps (second network).
It further decreases but, opposed to the other measures, the bimodality indicator reaches a stable level at around 4000 to 5000 steps.
Notice that the value at which it settles is larger than zero. 
This is due to the fact that not all agents develop the same extreme level of conviction but some (i.e. those at the interface between different opinion regions) remain slightly less convinced.
The saturation of the bimodality index at a low level therefore indicates that from time four to five thousand on there is a low but constant number of agents in an intermediate regime of conviction (cf. \cite{DiMaggio1996have}, p. 694).

\begin{figure}[ht]
\centering
\includegraphics[width=0.9\linewidth]{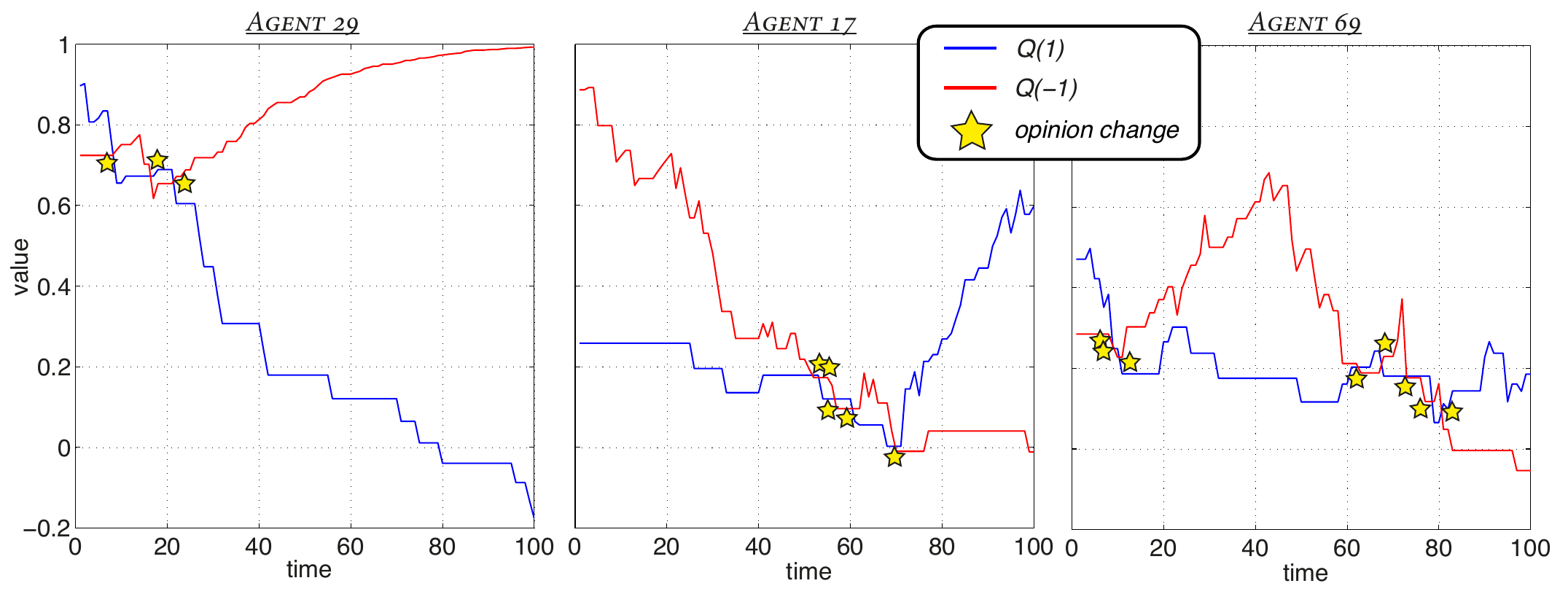}
\caption{Time evolution of three selected individuals in the above example run. The plot shows 10000 (on average 100 expressions by each agent). The yellow stars represent opinion change events where the values associated to the two alternatives change ranking.}
\label{fig:individuals01}
\end{figure}

To fully understand the dynamical behavior of the model, let us look at the temporal evolution of some particular agents.
In Fig. \ref{fig:individuals01} the values $Q_i(1)$ (blue) and $Q_i(-1)$ (red) are shown for three different agents.
Events where the ranking of values changes so that another opinion will be chosen in expression and response are highlighted by yellow stars. 
Aside from providing a better intuition of typical agent behaviors under the social feedback dynamics, this close-up view highlights the differences of the model with respect to continuous social influence models and physics-inspired binary state models (see Section \ref{sec:2}).

The first one (agent 29, located in the upper right corner of the networks in Fig. \ref{fig:BS01}) starts in a state where $Q(1) > Q(-1)$ but both values are at a relatively high level.
Initially, this agent is surrounded by an equal number of blue and red agents which leads to an decrease in both $Q(1)$ and $Q(-1)$ as the expected reward (local feedback) is zero in such a situation.
However, already at step 2500 (see second network in Fig. \ref{fig:BS01}) the neighborhood of 29 has aligned to -1 (red) and from that time on the value $Q(-1)$ increases steadily until it saturates at $Q(-1) = 1$ which is the expected social reward in an homogeneous surrounding.
Moreover, $Q(1)$ decreases step by step as the result of negative feedback on expressing $1$ due to the exploration probability ($\epsilon = 0.1$). 
$Q(1)$ will eventually approach -1 

This behavior is characteristic for many agents in the simulation without a too strong initial preference.
There is a initial phase of alignment in which a local consensus emerges -- weakly supported, at first -- which, once established, leads to an reinforcement of the respective view.
Notice that agents with a strong or moderate initial support of red in that cluster do in fact never change opinion.
Their values both tend to zero as well but the positive reinforcement of their initial preference begins before a change of ranking takes place.

Secondly, we consider agent 17 with a relatively strong initial preference for -1.
Her neighborhood remains unaligned for a rather long time and still is at step 5000.
In other words, the agent is located in the interface region between different opinion clusters that have emerged early in the simulation.
As shown in Fig. \ref{fig:BS01}, the agent forms part of a small cluster (along with 62, 75, 76) that is only slowly >>invaded<< by the blue opinion.
Around step 7000, however, most of her neighbors express preference for opinion 1 which leads to a rapid increase of $Q(1)$ in subsequent steps.

Finally, another node at the interface between two opinion regions is considered, namely agent 69.
With an initial preference for blue, this agent first aligns with the red cluster to which agent 29 belongs and learns to support this opinion rather strongly.
However, as the neighboring cluster around agent 17 aligns more and more on 1, expressing -1 yields a negative feedback with higher probability and leads to an decrease of $Q(-1)$.
Consequently, in between step 6000 to 8000 agent 69 experiences a second period of opinion changes the result of which is a slight preference of 1.
It is likely that a weak to moderate preference for blue will be stable as the agent is surrounded by 4 red and six blue agents.
However, there may be chains of random events that take the agent to temporarily supporting -1 again.

An important characteristic visible in this microscopic perspective is that agents do not immediately adapt their expressed opinion as a response to interaction with unaligned peers.
The imitation process implemented in many physics-inspired models of binary opinion dynamics is replaced by a continuous re-evaluation of the two options.
Periods of >>indecision<< during which individuals try different options alternate with periods during which agents show a clear preference of one or the other opinion.

Yet, this continuous adaptation is different from continuous models for opinion dynamics where an >>average opinion<< emerges within different network clusters.
To the contrary, the social feedback mechanism produces polarization by reinforcement of the support assigned to an opinion once a community is locally aligned.
In other words, the contraction mechanism leading to different opinion clusters in bounded confidence models (at least if the initial diversity is large enough) is replaced by a mechanism that drives opinion clusters to a higher degree of polarization (even if initial convictions are relatively close).



\subsection{Global Connectivity and Consensus Probability}

Previous work by \cite{Flache2011small} has started to address the effect of network density and long-range ties on the polarization process generated by different microscopic assumptions.
In their paper a model with positive social influence and selection based on homophily has been compared with a model where negative ties may form if opinions are too different and drive opinions still further apart.
While increased global connectivity due to the introduction of long-range ties fosters integration and consensus in the former, it fosters polarization under the latter assumptions.

\begin{figure}[ht]
\centering
\includegraphics[width=0.85\linewidth]{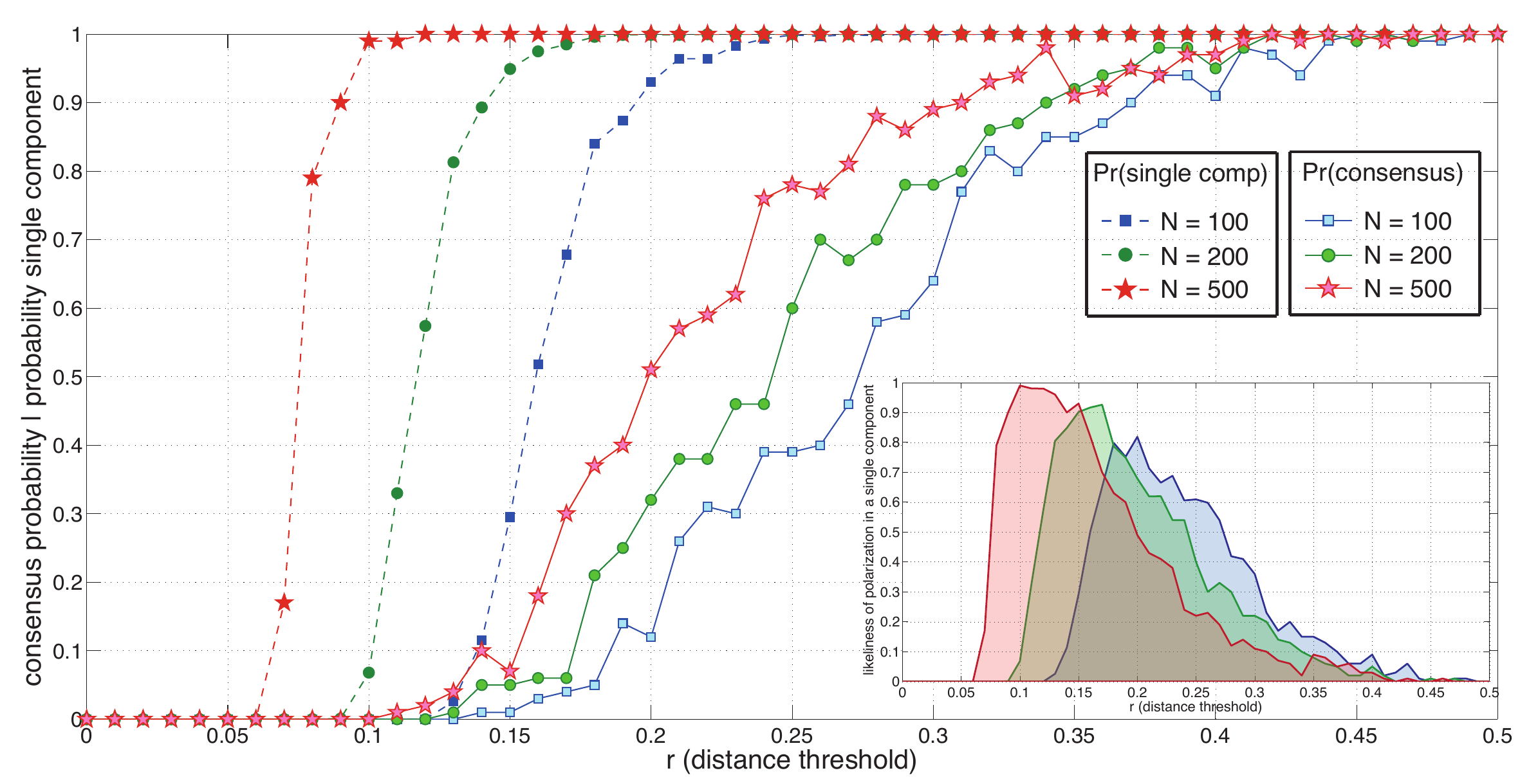}
\caption{Probability of consensus (all agents maximally support the same opinion) and of a single connected component as a function of the distance threshold $r$ governing global connectivity. The inset shows the respective likeliness that a graph with single connected component is generated and polarization observed on it.}
\label{fig:ConsensusProbability}
\end{figure}

In random geometric graphs, global connectivity is modulated by the distance threshold $r$.
If the radius $r$ is very small -- that is, below the critical threshold of $r_c = 4.52 / N$ \citep{Dall2002random} -- the network consists of many disconnected components.
As $r$ increases above the critical value ($r > r_c$), a giant connected component forms and the network becomes globally connected.
In this experiment we aim to provide a more complete picture of the model behavior on random geometric graphs by looking at its long-run behavior as a function of the radius $r$. 
For this purpose, we consider the consensus probability (the fraction of realizations that end up with all agents in the same opinion state) and compare it to the fraction of graph realizations that consist of a single connected component.
Notice that the latter is a very conservative measure of global connectivity because it considers graphs with a single isolated node as not connected.
Fig. \ref{fig:ConsensusProbability} shows this for three different system sizes of 100, 200 and 500 agents.
Each data point is obtained by averaging over 100 realizations of the model after $20000 \times N$ time steps.

The figure shows that polarization is very likely in a considerable range of networks in which all agents influence one another through direct or indirect influence paths.
There are three different regimes.
For very low connectivity, the networks consist of disconnected parts each performing an independent social feedback process.
This is the only regime where 
\revision{repeated averaging}
\citep{Abelson1964mathematical,Friedkin1990social,Friedkin1999choice} would predict persistent disagreement.
In the second regime, starting with an $r$ that decreases with the system size, the probability that the graph is connected sharply increases.
Notice that this transition takes place at values well above $r_c$ due to the fact that the probability of a single connected component is sensitive to single isolated nodes.
The consensus probability, on the other hand, only gradually increases and reaches one at a level of connectivity that is considerably higher compared to the level at which a single connected component becomes certain.
This means that in this region of the parameter space a large proportion of model realizations converges to polarized situation despite the fact that there is an influence path between all pairs of agents.
This observation becomes more pronounced when the number of agents increases.
For $N = 500$ and $r$ from $0.1$ to $0.15$, for instance, almost all realizations lead to a connected graph and a polarizing opinion formation process on it.
The inset of Fig. \ref{fig:ConsensusProbability} illustrates this by multiplying the probability of polarization and that of a single connected component.
Finally, as $r$ grows very large, the resulting networks become rather dense, less and less modular in terms of spatially divided subgroups, and consensus (all agents maximally support the same opinion) becomes the most likely outcome of the model.

\begin{figure}[ht]
\centering
\includegraphics[width=0.5\linewidth]{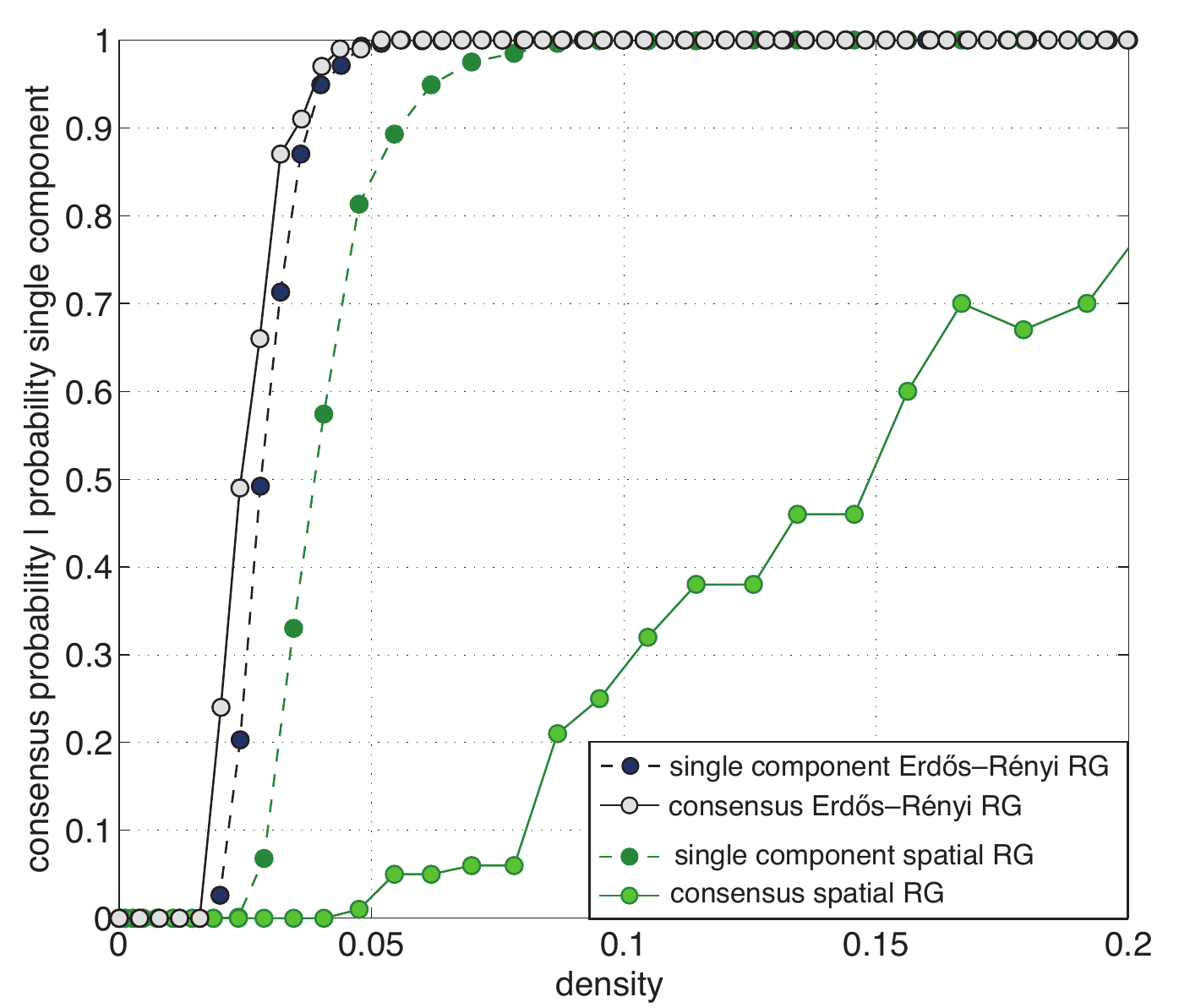}
\caption{Comparison of the random geometric graph (green) and the Erdős–Rényi random graph (black) for $N = 200$. Probability of consensus (all agents maximally support the same opinion) and a single connected component as a function of the network density.}
\label{fig:ComparisonRG}
\end{figure}


In order to show that the spatial organization into relatively (but not completely) segregated communities of agents (see the networks in Fig. \ref{fig:BS01}) is decisive for polarization to emerge, Fig. \ref{fig:ComparisonRG} reprints the results for $N = 200$ and compares them to an Erdős–Rényi random graph of the same size, that is, a graph without any spatial organization.
Notice that the probabilities are shown as a function of network density in order to compare the two models.
While the probability of consensus remains low compared to the probability of a single component in the spatial random graph model, both behave very similar in the Erdős–Rényi model.
That is, in the Erdős–Rényi graph we observe consensus whenever there is a single connected component.\footnote{Notice that the consensus probability is even slightly higher than the probability of a connected graph. The reason for this is that agents in different disconnected components may end up supporting the same opinion with a relatively high probability. For instance, in a random graph with two components and two opinions all agents converge to the same opinion in one half of the cases.}
This indicates that the spatial organization into several communities of agents that influence one another and interact less frequently with other groups is decisive for polarization to emerge under the social feedback process developed in the paper.
It points at the community structure of a network as the most important factor to explain polarization in the context of our model.
In the remainder of this paper, we will support this interpretation by game-theoretic considerations.

\section{Mathematical Characterizations}
\label{sec:math}

By treating opinion expression as an action and the social feedback as the payoff that an expression receives the proposed opinion model is strongly rooted in a form of reinforcement learning known as Q-learning \citep{Sutton1998reinforcement,Busoniu2008comprehensive}.
One of the interesting properties of this learning scheme is that its estimates of the value of different actions converge to the true expected utilities under certain conditions.
The model proposed in this paper can be seen as a repeated stochastic game played on a network, and therefore its rootedness in Q-learning provides a productive connection to previous work on games on networks (see \cite{Jackson2014games} and references therein).

In fact, coordination games -- as specified by our reward function (\ref{eq:reward})  -- have received particular attention and structural conditions for the stability of non-consensus configurations have been identified \citep{Morris2000contagion}.
These conditions are based on a notion of cohesion introduced in \cite{Wasserman1994social} and basically state that at least two subgroups with more in-group than out-group connections must exist in a network to allow for an equilibrium in which different actions may survive.

In the context of our model, the existence of multiple cohesive groups leads to group polarization processes that may take different directions within the different groups.
Using game-theoretic arguments, in this section, we show that the model leads to such a group polarization process in densely connected groups and that structural holes prevent a single opinion to spread over the entire network.
We use simulations to establish the relevance of these theoretical results for our model.
Moreover, we show that the interplay of these two essential properties leads to opinion bi-polarization in the social feedback model introduced in this paper.


\subsection{Decision Problem from the Individual Perspective}

Let us first look at a single agent in a fixed environment.
Assume therefore that a single agent $i$ is surrounded $k$ neighbors and denote by 
\begin{equation}
o_{\N(i)} = \frac{1}{k} \sum\limits_{j \in \N(i)}^{} o_j
\end{equation}
as the average expressed opinion in $i$'s neighborhood $\N(i)$.
The expected reward $i$ obtains on expressing $1$ is then given by $o_{\N(i)}$ and expressing $-1$ yields $-o_{\N(i)}$ on average.
Consequently, using these expectations, the update of the values $Q_i(1)$ and $Q_i(-1)$ is given by
\begin{eqnarray}
\begin{array}{l l}
Q_i^{t+1}(1) & = Q_i^{t}(1) + \Pr(o_i^t = 1) \alpha (o_{\N(i)} - Q_i^{t}(1))\\
Q_i^{t+1}(-1) & = Q_i^{t}(-1) + \Pr(o_i^t = -1) \alpha (-o_{\N(i)} - Q_i^{t}(-1))
\end{array}
\label{eq:VSingleAgent}
\end{eqnarray}
where $\Pr(o_i^t = o)$ are the probabilities that $i$ performs the respective action (expresses $1$ or $-1$) at time $t$. 
In the model this depends on the current value of $Q_i^t(1)$ and $Q_i^t(-1)$, but we just notice that if we allow for exploration ($\epsilon > 0$, whatever small) these probabilities are strictly positive.
Under this assumption, therefore, the fixed point of (\ref{eq:VSingleAgent}) is given by $Q_i^*(1) = o_{\N(i)}$ and $Q_i^*(-1) = -o_{\N(i)}$ and the respective value difference by $\Delta Q_i^* = 2 o_{\N(i)}$.

Therefore, if an individual is in a homogeneous environment with all neighbors in the same state (i.e., $o_{\N(i)} = 1$ or $-1$) $i$'s conviction settles at a maximal value in compliance with neighbors whatever $i$'s initial assignment of values has been.
That is, $i$ becomes maximally convinced of the respective opinion.
Noteworthy, as only the expressed opinions of neighbors are visible to $i$ this occurs even if neighbors are only weakly supportive of that view.
Hence, in a homogeneous environment the model gives rise to a radicalization process by which weakly convinced members approach maximal conviction irrespective of the initial conviction in the neighborhood.

In this sense, the model captures a dynamical process of group polarization by which a group initially inclined into an attitudinal direction delivers a more extreme judgement after discussion (see \cite{Sunstein2002law} and references therein as well as \cite{Lord1979biased} on biased assimilation).
Contrary to the limited argument pool or biased processing argument, however, our model reveals a social influence route to polarization where agents explore the social acceptability of their opinion and conviction is strengthened in the absence of opposing voices.


\subsection{Gate Keeping at Structural Holes}

\begin{wrapfigure}{r}{0.39\linewidth}
\includegraphics[width=0.99\linewidth]{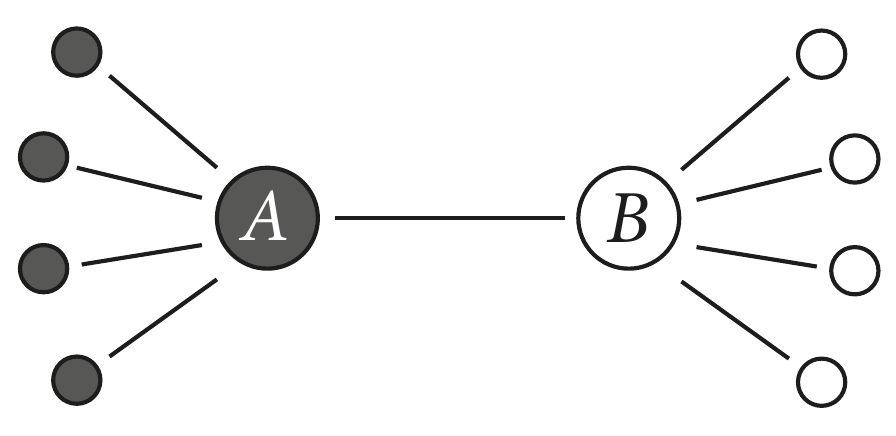}
\caption{Interaction of two agents $A$ and $B$ who are each linked to an opinion community of different sign.}
\label{fig:StructuralHole01}
\end{wrapfigure}
Real social networks are not complete.
They are rather sparse and exhibit complex structure. 
They tend to come in clusters of tightly connected groups within which information flows rapidly, but across which influence and information flow is reduced.
Ties connecting different clusters (along with the respective nodes) have received particular attention in social network analysis \citep{Burt2004structural,Borgatti2009network} due to their gate keeping role and strategic position which may come as a competitive advantage in terms of access to information, but also as a challenge to maintain connections with conflicting groups.

In the context of opinion dynamics, it is important to understand whether and at what pace opinions spread across different communities.
For this purpose, let us consider the interaction of two agents $A$ and $B$ both connected to a different community homogeneously supporting different opinions (see Fig. \ref{fig:StructuralHole01}).
Notice that this is the paradigmatic example of a bridge as considered in \cite{Burt2004structural}.
Let us denote by $k_{A}$ the number of neighbors of $A$ and by $k_B$ the number of neighbors of $B$ respectively.
Assume that the opinion in $A$'s community is -1 ($o_{\N(A)\setminus B} = -1$) and that $B$'s community adheres to opinion 1 ($o_{\N(B)\setminus A} = 1$).
We leave the respective support levels in the two communities unspecified and only look at the best choices for $A$ and $B$.
From the point of view of an interaction between $A$ and $B$ we can render such a situation as a game with the following bi-matrix for the single shot game:\\
\begin{center}
{\def\arraystretch{2}\tabcolsep=10pt
\begin{tabular}{c|cc}
			& $-1$ 		& $1$ \\ \hline 
$-1$ 	& $(1 , \frac{1-k_B}{k_B+1})$ 	& ${\bf(\frac{k_A-1}{k_A+1} , \frac{k_B-1}{k_B+1})}$ \\
$1$ 	 & $(-1,-1)$ & $(\frac{1-k_A}{k_A+1},1)$\\
\end{tabular} 
}\\
\end{center}
It is immediately clear that as soon as the size of the communities linked to $A$ and $B$ respectively exceeds one, the Nash equilibrium of the game is given by $(o_A^* = -1 , o_B^* = 1)$ such that both agents maintain the opinion of their respective group.
This shows that different opinion clusters are a stable outcome of the opinion formation process on clustered interaction networks such as those considered in the previous section.



\begin{figure}[ht]
\centering
\includegraphics[width=0.5\linewidth]{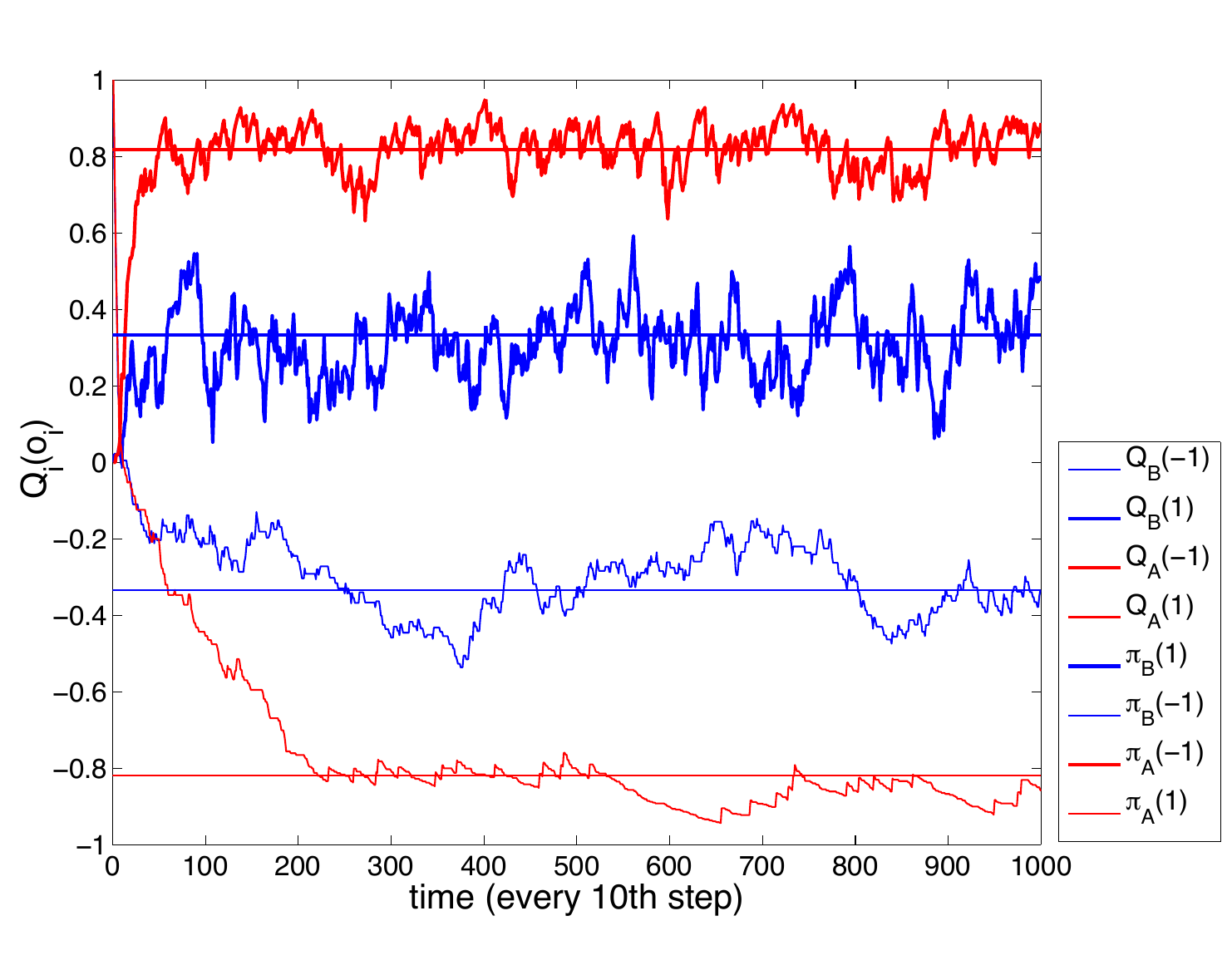}
\includegraphics[width=0.48\linewidth]{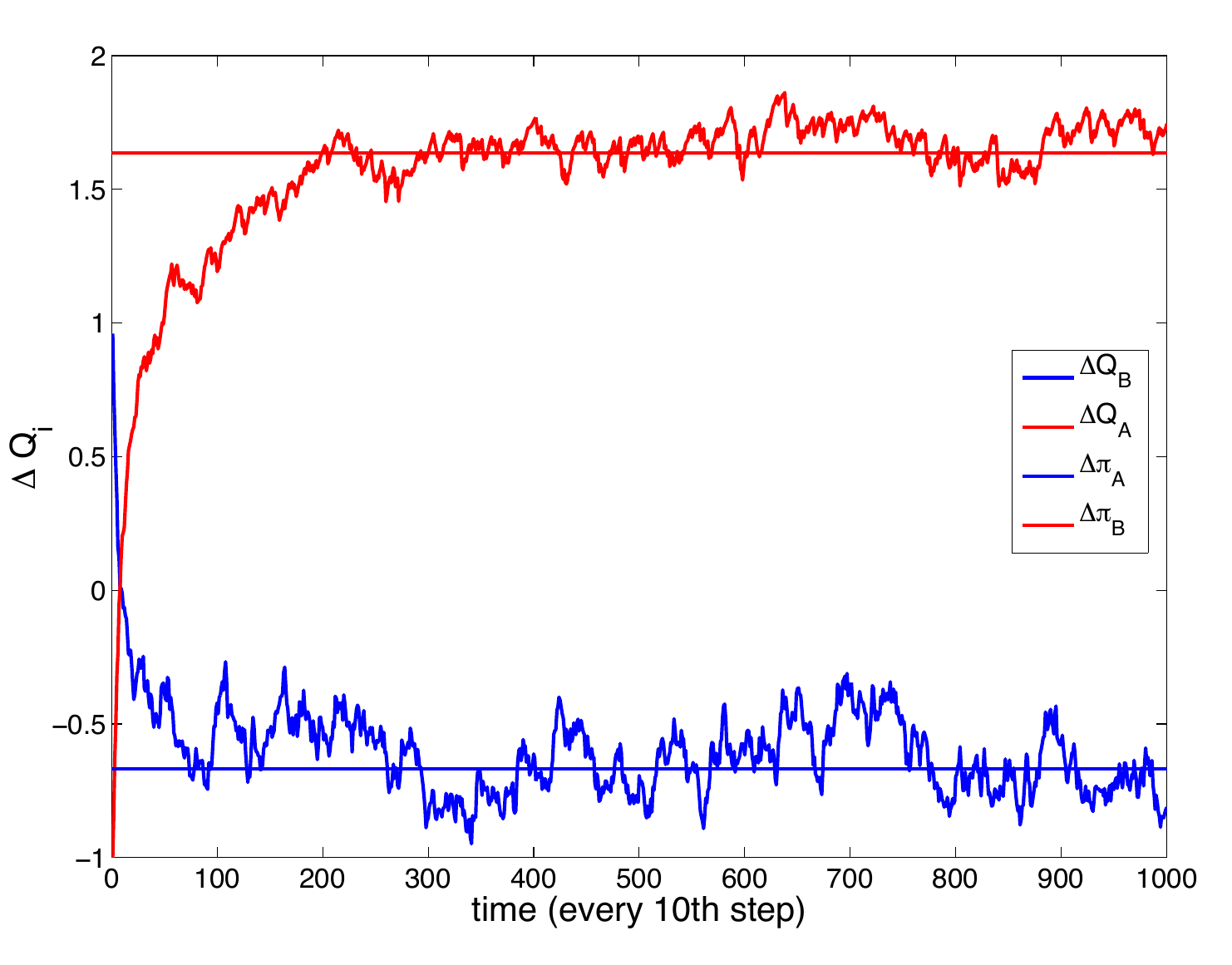}
\caption{Dynamical evolution of two agents A and B each linked to a fixed community of opposed sign. A is connected to $k_A = 10$ agents supporting -1, B to $k_B = 2$ agents supporting 1. A learning rate $\alpha = 0.02$ and exploration rate $\epsilon = 0.1$ is used. The figure shows a quick convergence of the $Q$-values to the payoffs of the associated opinion game.}
\label{fig:GateKeeping_10_2}
\end{figure}

In order to show that the game shown in the game matrix above can provide a valid approximation for the proposed social feedback model, let us briefly consider the relation between the above payoff matrix and the dynamics (\ref{eq:QUpdate}) of our model.
Therefore, assume that agent B forms a community with just two other agents ($k_B = 2$) and agent A is part of a bigger community with ten others ($k_A = 10$) and that A's community is in favor of $-1$ and B's community supports $1$.
Consequently, the theoretical payoffs for A are $\pi_A(1) = -9/11$ and $\pi_A(-1) = 9/11$ and for B $\pi_B(1) = 1/3$ and $\pi_B(-1) = -1/3$.
Fig. \ref{fig:GateKeeping_10_2} shows the evolution of the four values $Q_i(o_i)$ along with the payoffs as specified above (l.h.s.) and the resulting differences $\Delta Q_i$ and $\Delta \pi_i$ (r.h.s.).
The agents A and B are initialized such that they strongly disagree with their respective group.
However, both quickly approach the game payoff values such that, in effect, they learn to play the associated opinion game.

\subsection{Opinion Games and Cohesive Sets}

The previous analysis points out the relevance of game-theoretic arguments for the analysis of the convergence behavior of the proposed model and the identification of conditions for the emergence of polarization.
Coordination games and games of strategic complements more generally have received particular attention in the literature on games on networks \citep{Morris2000contagion,Jackson2010social,Jackson2014games}.
From this perspective, our model belongs to the category of semi-anonymous games \citep{Jackson2010social} in which only the number of neighbors taking the different actions (expressing their opinion) plays a role regarding the best response of a player and not the specific interaction partners. 
In other words, players assign equal importance to all of their local neighbors.

In the context of this type of games, a particular notion of group cohesion \citep{Wasserman1994social} has proven an effective tool to characterize structural conditions for contagious spreading of one action to the entire population and conversely also for the stable coexistence of both alternatives in the population \citep{Morris2000contagion}.
Given a network along with a pay-off structure such as given with (\ref{eq:reward}), the challenge is to identify action configurations over the network that are stable in the game-theoretic sense.
That is, configurations where no player independently would be better off by switching to the other alternative.
Notice that in our model, a players expression $o_i$ is a best response choice if at least half of its neighbors favor that opinion as well.
Cohesion measured as >>the relative frequency of ties among group members compared to non-members<< \citep[p. 64]{Morris2000contagion} is therefore a direct way to check whether a group of agents plays at equilibrium.
Namely, given a group $S$ of agents such that each member has more connections to agents in $S$ than to others (in the complementary set $\bar{S}$), then this group plays at equilibrium whenever all members choose the same action.
In consequence, whenever there are at least two distinct groups with cohesion greater or equal to $1/2$ (more in-group then out-group links) in a network, configurations in which different opinions are expressed in the different groups are stable.\footnote{Cf. \cite[Prop. 5]{Morris2000contagion} and \cite[Prop. 3.3]{Jackson2014games}.}

More formally and following \cite{Morris2000contagion}, let us define the fraction of neighbors of agent $i$ that are also in $S$ as
\begin{equation}
\pi(S|i) = \nicefrac{ \sum\limits_{j \in S}^{} a_{ij} }{\sum\limits_{j=1}^{N} a_{ij}},
\end{equation}
where $a_{ij}$ are the elements of the adjacency matrix.
The cohesion of the entire group $S$ of agents is then defined as the smallest value $\pi(S|i)$ of all the individuals in $S$, that is
\begin{equation}
coh(S) = \min_{i \in S} \pi(S|i).
\label{eq:cohesion}
\end{equation}

As mentioned above, the reward system (\ref{eq:reward}) used in this paper corresponds to a symmetric coordination game for which the relevant cohesiveness level is $1/2$.
For the stability of a non-consensus configuration it is necessary that there is at least one partition of the network into $S$ and its complement $\bar{S}$ such that both $S$ and $\bar{S}$ are at least $1/2$-cohesive.
The intuition behind that is relatively simple. 
If, for example, $coh(S) < 1/2$ this means that there is at least one agent $i \in S$ that has more neighbors in $\bar{S}$ than in $S$.
Therefore, if one opinion (say $1$) is expressed in $S$ and the other ($-1$) in $\bar{S}$, agent $i$ would improve its expected pay-off by switching to $-1$ destroying the group consensus in $S$.
As other agents in $S$ are connected to $i$ its switching may potentially lead to a cascade of further opinion changes by other members of the group $S$ and, depending on the connectivity structure, global alignment on expressing $-1$ may result. 

Let us finally illustrate the cohesion concept by a brief look at the example from the previous section where two agents are connected to different communities.
It is easy to compute that A's community $S_A$ with $|S_A| = 11$ is $10/11$-cohesive because $A$ has 11 neighbors 10 of which are in $S_A$ and the rest of the agents is only connected to $A$ leading to a cohesion value of one.
Respectively, $S_B$ with $|S_B| = 3$ is $2/3$ cohesive because $B$ has three neighbors and two of them in $S_B$.
Hence, the cohesion condition is satisfied for both subsets ($coh(S_A) > 1/2$ and $coh(S_B) > 1/2$) and consequently the polarized outcome is stable.

\subsection{Cohesion in Two-Community Graphs}

Let us generalize this example and consider the case of two communities with $M$ and $L$ agents where we assume that most connections are within and only a few across the two communities.
Let $p$ denote the probability of an (undirected) inter-community link between an agent in $S_M$ and an agent in $S_L$ and $(1-p)$ the probability for intra-community connections.
The obvious advantage of this graph model is that we (externally) define a partition of the network for which we may study how the stability of the non-consensus configurations is affected as $p$ increases.
Meanwhile this setting can be seen as a prototypic situation that occurs also in more complex social networks.

\begin{figure}[ht]
\centering
\includegraphics[width=0.6\linewidth]{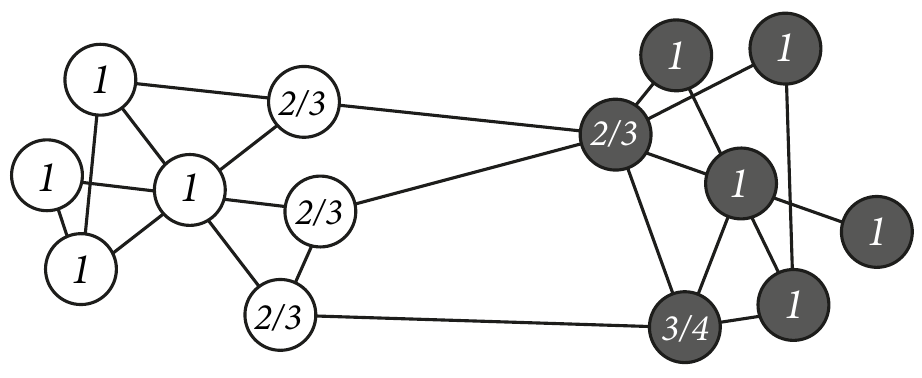}
\caption{A small two-community graph with two densely connected components and a few >>long-range<< ties connecting them. The fractional numbers shown on the nodes correspond to the respective fraction of neighbors with the same opinion $\pi(S | i)$.}
\label{fig:TwoComNetwork}
\end{figure}

In fact, there is a long tradition in studying the two-community topology in population genetics \citep{Wright1943isolation} and also theoretical work on opinion dynamics has relied on this as a stylized description of segregated communication structures \citep{Dandekar2013biased,Banisch2014microscopic,Lamarche2016information}.
One could think of it in spatial terms such as, for instance, as a set of villages with intensive interaction among people of the same village and some contact across, but it could also be related to homophily regarding to social classes, ethnicity, religious communities, etc.
Fig. \ref{fig:TwoComNetwork} shows a small realization of such a two-community setting.
There are 7 nodes in each community and three links that connect agents from different groups.
The numbers shown on the nodes represent the individual ratios $\pi(S | i)$ given that partition.
In this case, these values give rise to a cohesion of $coh(S) = \min_{i \in S} \pi(S|i) = 2/3$ for both subgroups rendering a bi-polar situation where two competing view coexist as stable.

The ratio $\pi$ of an agent in (say) community $S_M$ is given by
\begin{equation}
\pi(S_M | i) = \nicefrac{ \sum\limits_{j \in S_M}^{} a_{ij} }{\sum\limits_{j \in S_M \cup S_L}^{N} a_{ij}} = \nicefrac{m_i}{m_i + l_i}
\end{equation}
where $m_i$ ($l_i$) is used as a shorthand for the number of links that $i$ maintains to agents in $S_M$ ($S_L$).
Given that links are independently created with probability $p$ and $1-p$ respectively, each row of the adjacency matrix can be seen as the result of a finite sequence of Bernoulli trails.
Consequently, for an agent $i \in S_M$ the probability $\Pr_{in}(m)$ that exactly $m$ links to agents in the same community are created is given by
\begin{equation}
\Pr_{in}(m) = (1-p)^m p^{M-1-m} \binom{M-1}{m}
\end{equation}
where $M-1$ corresponds to the number of potential within-community links excluding self-connections.
The probability of exactly $l$ across-community connections $\Pr_{out}(l)$ reads
\begin{equation}
\Pr_{out}(l) = p^l (1-p)^{L-l} \binom{L}{l}.
\end{equation}
Configurations where different opinions are expressed in the different communities loose their stability if $m_i < l_i$ for at least one agent $i$, because $1/2$-cohesion of the entire community rests upon $1/2$-cohesiveness of each individual with respect to the group (\ref{eq:cohesion}).
For each single agent in $S_M$ the probability that (s)he is less than $1/2$-cohesive is given by
\begin{equation}
\Pr(m < l) = \sum\limits_{m=0}^{M-1} \Pr_{in}(m) \sum\limits_{l=m+1}^{L} \Pr_{out}(l) = q_M.
\end{equation}
For convenience we denote this probability as $q_M$.
Likewise, the probability to be less than $1/2$-cohesive for an agent in $S_L$ (denoted by $q_L$) is obtained by exchanging $m$ with $l$ and $M$ with $L$ respectively:
\begin{equation}
\Pr(l < m) = \sum\limits_{l=0}^{L-1} \Pr_{in}(l) \sum\limits_{m=l+1}^{M} \Pr_{out}(m) = q_L.
\end{equation}
For a network consisting of $M+L$ agents, the probability that there is no agent for which $m_i < l_i$ can be obtained by a similar reasoning conceiving the problem as a sequence of $M+L$ Bernoulli trails with probabilities $q_M$ and $q_L$ respectively.
Following this argument, the probability that no agent in $S_M$ is less than $1/2$-cohesive is given by $(1-q_M)^M$ and for $S_L$ it reads $(1-q_L)^L$.
Therefore, the probability that the partitions $S_M$ and $S_L$ both satisfy the cohesion condition for the stability of different actions in the different communities is given by
\begin{equation}
\Pr[coh(S_M) \geq 1/2 \wedge coh(S_L) \geq 1/2 ] = (1-q_M)^M (1-q_L)^L.
\label{eq:PrCohesive}
\end{equation}

\begin{figure}[ht]
\centering
\includegraphics[width=0.6\linewidth]{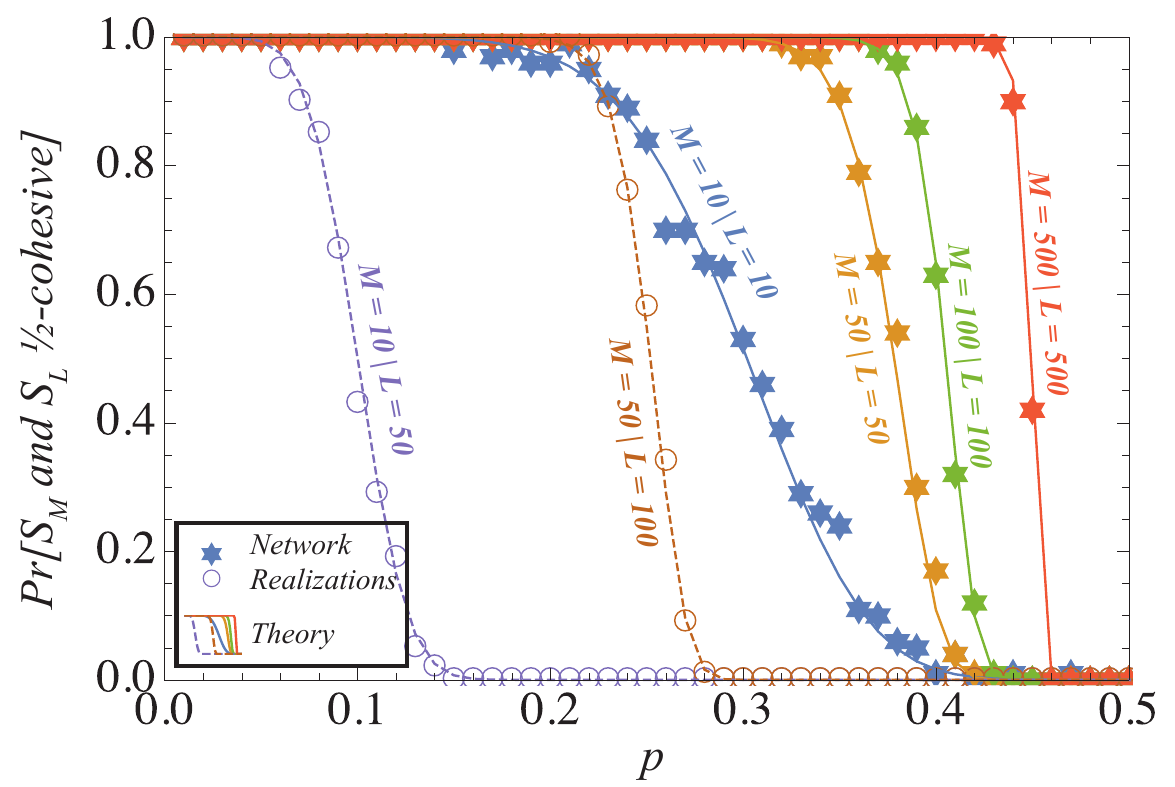}
\caption{ Probability that both communities are at least $1/2$ cohesive as a function of $p$. The figure compares the theoretical results (\ref{eq:PrCohesive}) with an average over 100 network realizations.}
\label{fig:PrCohesive}
\end{figure}

Fig. \ref{fig:PrCohesive} compares the theoretical probability  (\ref{eq:PrCohesive}) that both communities are at least $1/2$-cohesive with an average over 100 network realizations and demonstrates the correctness of the previous calculations.
It shows four symmetric ($M = L$, solid stars) and two asymmetric cases ($M \neq L$, dashed circles), but we shall focus on the symmetric ones here.
In the cases where $M = L$ we observe that the rewiring probability $p$ at which the transition from stability to instability occurs increases with the number of agents in the communities.
At the same time it becomes sharper with increasing size which indicates a discontinuous phase transition in the limit of an infinite system.
In fact, it is easy to see that when $M \rightarrow \infty$, $L \rightarrow \infty$ and $M = L$ the fraction of intra- and inter-community links is precisely $(1-p)$ such that the critical value at which the communities loose $1/2$-cohesion is precisely $p^* = 1/2$.

\begin{figure}[ht]
\centering
\includegraphics[width=0.49\linewidth]{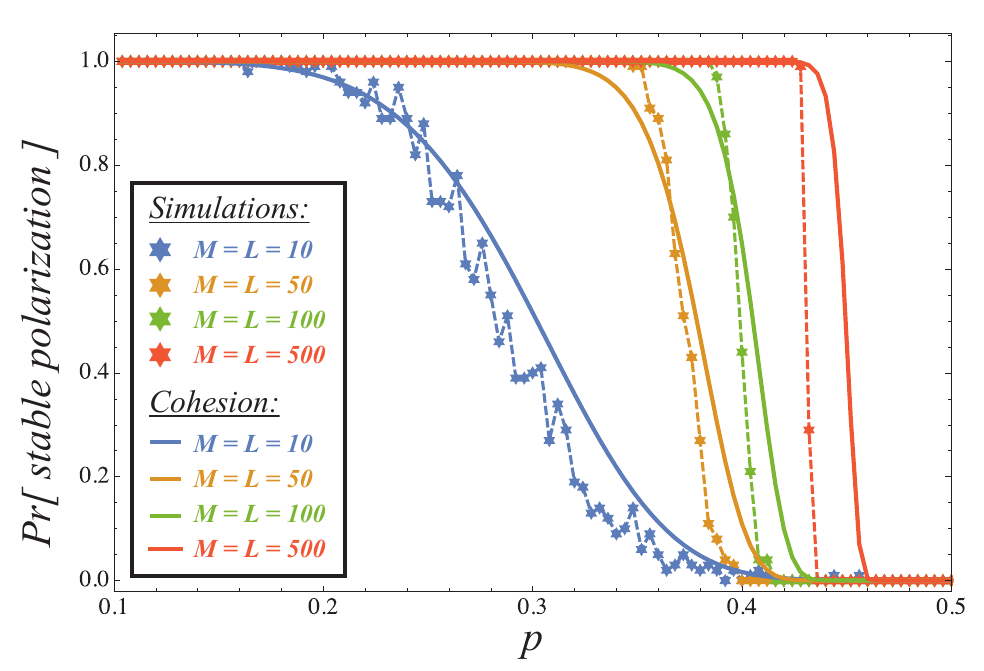}
\includegraphics[width=0.49\linewidth]{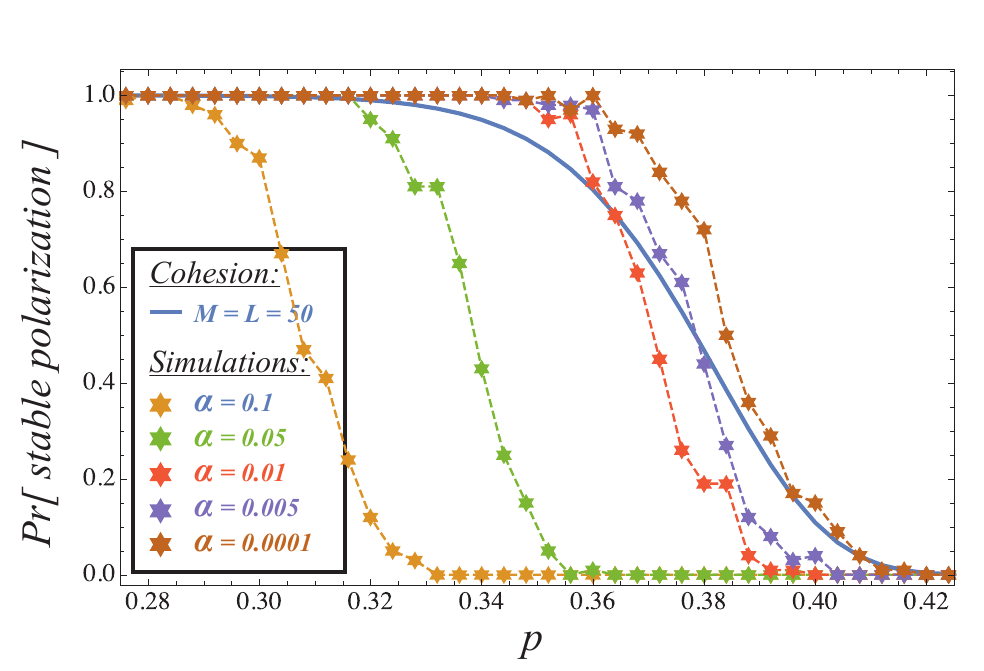}
\caption{Probability of persistent polarization in a suite of 100 simulations per data point (inter-community coupling $p$) is compared to the $1/2$-cohesion probability as computed by (\ref{eq:PrCohesive}). On the l.h.s. the comparison is shown for various community sizes from 10 to 500 with a fixed learning rate $\alpha = 0.01$. On the r.h.s. the influence of the learning rate is studied for a system of fixed size $M = L = 50$.}
\label{fig:PrSimulations}
\end{figure}

The combinatorial analysis of cohesion hence predicts a phase transition from stable bi-polarization to consensus on the two island graph which becomes sharper with increasing system size.
In order to assess the significance of this result for the convergence behavior of the model, we initialize the two--island system in a state of maximal polarization and check if polarization persists under the opinion reinforcement mechanism.
That is, we initialize agents in community $S_M$ with a high support for opinion $-1$ by setting $Q^0_{i \in S_M}(-1) = 1$ and $Q^0_{i \in S_M}(1) = -1$ such that $\Delta Q^0_{i \in S_M} = -2$ and agents in the other community just the opposite such that $\Delta Q^0_{i \in S_L} = 2$.
For different $p$, the system is then iterated for a relatively long period of $20000 \times N$ steps ($N = M + L$) and we compute the fraction of realizations out of 100 that reached consensus at that time.
Consensus (i.e. all agents express either $1$ or $-1$) is a suitable indicator because once reached the respective opinion is globally reinforced and the probability that one agent reverses due to finite-size fluctuations is effectively zero.
In a first experiment the community sizes are varied from $10$ to $500$ (Fig. \ref{fig:PrSimulations}, l.h.s) and the learning rate is set to $\alpha = 0.01$.
In a second experiment the size is fixed to $M=L=50$ and the influence of the learning rate $\alpha$ is analyzed (reported on the r.h.s of Fig. \ref{fig:PrSimulations}).

The l.h.s of Fig. \ref{fig:PrSimulations} shows the results of this experiment for different system sizes and compares them to the theoretical 1/2-cohesion probability (\ref{eq:PrCohesive}).
All in all, a relatively good agreement between the simulations and the theoretical curves is observed which shows that the combinatorial analysis of group cohesion provides an accurate prediction for the persistence of polarization in the two-community scenario.
However, there are two effects that the cohesion probability does not properly account for.
First, for the large system with $M = L = 500$ the transition from stable polarization to consensus takes place at a lower value of $p$ compared to the theoretical curve.
Second, the transition observed for the model dynamics is generally sharper than the theoretical prediction.
In fact, the transition is rather sharp already for a relatively small system of 100 agents.

The r.h.s of Fig. \ref{fig:PrSimulations} -- looking at the influence of the learning rate $\alpha$ -- sheds some light on both deviations.
First, regarding the lower critical value for large systems it becomes clear that the learning rate must be small enough to match with the theoretical transition point.
For a system of 100 agents a learning rate of $\alpha = 0.05$ still leads to a transition from polarization to consensus at a significantly lower value of inter-community coupling whereas with $\alpha < 0.01$ the theoretical curve is approached well.
As the learning rate governs the fluctuations of the $Q$-values, these results suggest that the closer the community graph is to a group cohesion of $1/2$, the more likely it becomes that one community is invaded due to a sequence of out-group interactions by which agents in one community occasionally express the opinion of the other enabling that opinion to spread throughout the entire cluster.
The probability of such a cascading invasion increases with the size of the system so that the theoretical curves are matched only if the learning rate (and hence the fluctuations in the $Q_i$) is reduced.

Cohesion is a structural measure which characterizes stable network configurations in the following sense: given the configuration of all players' actions no agent alone can improve its payoff by switching to the opposed action.
It thus assumes that all players know their best response and play accordingly.
In the opinion model, however, agents learn which opinion is favored in their local environment, and they do so in a sequential manner receiving feedback only from one peer at a time.
This means that they may deviate from >>best response behavior<< from time to time due to imperfect estimates $Q$ of their expected reward.
A perfect matching can thus be observed only in the limit $\alpha \rightarrow 0$. 

The second observation that the transition is generally sharper under learning dynamics has two distinct reasons.
On the one hand, notice in Fig. \ref{fig:PrSimulations} (r.h.s.) that the match between theory and simulations becomes fairly accurate in the lower part ($p > 0.39$) of the curve with decreasing $\alpha$ (see brown stars for $\alpha = 0.0001$).
In this parameter regime where cohesion is generally very close to 1/2 a finite learning rate $\alpha > 0$ may lead to a cascade triggered by fluctuations in the $Q_i$ as described above.
On the other hand, for $p < 0.39$ we notice a higher probability of persistent opinion polarization -- even increasing with decreasing $\alpha$ -- which points at an aspect the cohesion probability (\ref{eq:PrCohesive}) does not account for.
The reason for this is that the theoretical probability (\ref{eq:PrCohesive}) is computed by fixing the network partition to $S_M$ and $S_L$.
In the random assignment of connections, it may, however, happen that a single node becomes relatively disconnected from its predefined set and at the same time cohesively connected to the other set.
While (\ref{eq:PrCohesive}) would predict a loss of cohesion of $S_M$ or $S_L$ the entire network still possesses two cohesive sets (slightly different from $S_M$ and $S_L$) on which different opinion are stably expressed.

\section{Conclusion}
\label{sec:conclusion}

This paper makes the following main contributions:
\begin{enumerate}
\item 
It develops a new mechanism for the emergence and persistence of opinion bi-polarization which is based on a parsimonious set of assumptions.
As opposed to previous models aiming at an explanation of bimodal opinion distributions, the proposed social feedback model does not rely on negative social influence \citep{Mark2003culture, Macy2003polarization, Baldassarri2007dynamics, Flache2011small} or assumptions about opinion homophily or bounded confidence \citep{Axelrod1997dissemination,Deffuant2000mixing,Hegselmann2002opinion,Maes2013differentiation,Duggins2017psycologically}.
It discriminates an internal conviction and an externally expressed opinion by a learning process which reinforces an agents private conviction in its expressed opinion based on the rewarding or non-rewarding experiences made by communicating ones views in a social neighborhood.
In large social networks that consist of different communities this social feedback mechanism gives rise to group polarization processes by which members of the same community become collectively more convinced of one opinion.
This process plays out independently in different communities even if weak ties and individual links are maintained across the different groups because the rewards gained by adopting and expressing the group opinion is larger than the rewards attainable in less frequent interactions with the out-group.
As a consequence, for two connected individuals that are members of different groups it is desirable to maintain their respective group opinion because the rewarding experiences from communication within their group outweigh the negative experiences of disagreement when they occasionally encounter.
\item
The paper introduces reinforcement learning to opinion dynamics modeling.
While reinforcement learning as a basic model of social behavior is not new and is, for instance, at the core of Homans' work on social exchange theory \citep{Homans1958social,Homans1974social}, its application and interpretation as a model for opinion dynamics and polarization in particular is.
In our model, we consider that opinion expression is a decision problem that involves an internal evaluation of the expected effect of available expression alternatives based on the social feedback these expressions previously received.
Rejection and confirmation of expressed opinions by peers lead to a re-evaluation of the different options which is mediated via a reward signal.
This mechanism leads to the formation of strong convictions in an opinion -- that is, a clear evaluation that one opinion is preferred over the other -- based on how acceptable it is to advocate that standpoint in a social neighborhood.
It is noteworthy that studies in neurobiology suggest that >>social influence mediates very basic value signals in known reinforcement circuity<< \citep{Campbell2010opinion}.
However, while neurobiology posits that the >>rewarding properties of social behavior may have evolved to facilitate group cohesion and cooperation<< \citep{Ruff2014neurobiology}, our model suggests that polarization (as opposed to cohesion) across groups may be a side-effect of these rewarding properties.
In other words, human ability to coordinate with in-groups comes at the expense of a likely alienation to out-groups, which we could refer to as a >>tragedy of coordination<<.
\item
Through the use of reinforcement learning, the paper establishes a link between models of opinion formation and standard game-theoretic notions of equilibrium and explores its usefulness in the analysis of the convergence behavior of the model.
A particular advantage of social influence network theory \citep{French1956formal,Abelson1964mathematical,Friedkin1990social,Friedkin1999choice,Friedkin2016network,Parsegov2016novel} is analytical tractability 
\thirdrevision{Some extensions of the theory prompted by the polarization problem}
introduce a non-linearity that is difficult to handle with analytical tools \citep{Hegselmann2002opinion}.
The model we put forth can also be seen as a non-linear extension to these models which, while being plausible and psychologically well-justified, provides a powerful tool to establish the connection to game-theory and well-established equilibrium concepts.
Namely, by the Q-learning scheme adopted to operationalize the opinion formation process agents learn to associate values to the different opinion expressions that converge to the payoffs of the corresponding >>opinion game<<.
This has been shown and used throughout Section \ref{sec:math}.
The theory of games on networks in particular \citep{Morris2000contagion, Jackson2014games} has proven useful to establish conditions under which persistent disagreement can be expected by the mechanism proposed in this paper.
\item
In that context, the notion of cohesive sets \citep{Morris2000contagion} 
has been shown to provide a useful structural measure for the characterization of network conditions for polarization.
In particular, the paper shows that a bimodal distribution of opinions is a stable outcome of the social feedback model of opinion dynamics whenever at least two communities exist in a network with more connection within than across groups.
In such a situation the reinforcement mechanism gives rise to a group polarization process that may take different directions within the different groups. 
That is, the cohesive structure of a network is decisive for bi-polarization in the context of our model and the effect of different connectivity patterns on the model dynamics (such as, for instance, long-range ties, \cite{Centola2007complex, Flache2011small}) consequently becomes a question of whether the cohesive structure of the network undergoes a qualitative change or not.
\item
In this sense, our model shows that persistent, even increasing polarization may be obtained despite a significant exposure to attitude-challenging content \citep{Bakshy2015exposure}.
Moderate levels of homophily and network segregation are sufficient for opinion polarization.
\end{enumerate}

This paper paves the way for quite a few interesting topics to be addressed by future research.
For instance, the model can be seen as an abstraction of recently proposed persuasion models of polarization in which opinion--confirming interactions reinforce commitment to that view whereas disconfirming interaction weakens opinion support due to biased argument processing \citep{Dandekar2013biased} or biased argument pools \citep{Sunstein2002law,Maes2013differentiation}.
On the other hand, the model is based on a more simple reinforcement procedure which does not assume any more complex kind of argument processing on the side of the agents but instead conceives opinion evaluation and consolidation as the result of positive and negative experiences that are assumingly related to agreement and disagreement with peers.
Despite these different interpretations, the collective dynamics of our social feedback model and the argument persuasion model proposed in \cite{Maes2013differentiation} share certain similarities.
In fact, we conjecture that the incorporation of an opinion homophily mechanism at the level of convictions in our model would be compatible with the homophily concept employed in the model by \cite{Maes2013differentiation} and that the learning rate of the reinforcement scheme used in this paper can be adjusted to the impact that the adoption of a new argument has in their model.


Second, the random geometric graph model used in the first part of the paper has mainly been chosen to illustrate the possibility of stable bi-polarization in a connected graph.
On the other hand, the main purpose of the two--community setting used in Section \ref{sec:math} has been to show that the game-theoretic characterization of non-consensus equilibria in terms of cohesive sets provides a valid characterization for the opinion model as well.
Future work should explore more complex and more realistic social networks.
In particular, our analysis points at the relevance of community structure of a graph which, in network science, is typically captured by measures of modularity \citep{Girvan2002community,Fortunato2010community}.
Future work should clarify on mathematical grounds the relation between the notion of cohesion as used in the game--theoretic literature and different conceptions of modularity with the respective approaches to community detection.
In addition, we envision that the proposed model is  actually well--suited as a tool for community detection and the identification of structural holes in real social networks.

Real social networks are not static but evolve in time and may undergo structural changes.
A natural model extension is therefore to provide agents with the possibility to search for new interaction partners if their opinion deviates from the opinion in their current neighborhood.
In the framework of reinforcement learning that we propose this can be incorporated in different ways.
For instance, given their current opinion agents could learn the rewards they can expect from each other agent in the population and base their decisions of whom to interact with on that evaluation.
In the current setting of binary opinions this would either lead to consensus with all agents strongly supporting one opinion or to a complete bi-separation of the network into two groups strongly supporting different views.
Other outcomes can be achieved by making network changes costly \citep{Bojanowski2011coordination}.

On the whole, we believe that linking opinion dynamics with social feedback mechanisms bears great potential for modeling opinion formation processes in different social settings and the model presented here should be understood as an initial implementation of this paradigm.
The theory of reinforcement learning in which different behavioral options (opinion expressions in our case) are constantly re-evaluated based on the rewards they give in a certain environment provides a theoretically convenient and very general framework to address a wide variety of questions that are relevant in the science of opinion dynamics.
Most notably, it shifts the explanatory focus from mechanism of social influence and opinion exchange to the incentives and rewards of opinion expression in different social settings which is becoming more and more germane to understanding opinion exchange processes in social media platforms.

\section*{Acknowledgment}

This project has received funding from the European Union’s Horizon 2020 research and innovation programme under grant agreement No 732942 (www.\textsc{Odycceus}.eu).
We are grateful for the repeated discussion of the ideas described in the paper by the members of \textsc{Odycceus} and Sharwin Rezagholi in particular.
The feedback by participants of the >>Interdisciplinary Workshop on
Opinion Dynamics and Collective Decision<<, July 5-7, 2017, Bremen, Germany and the >>Social Simulation Conference (SSC) 2017<<,  Dublin, Ireland, is also gratefully acknowledged.
We also acknowledge a final reading by Michael Mäs.
Finally, the paper substantially improved due to the comments of three anonymous referees.

\small
\bibliographystyle{apalike}

\end{document}